\documentclass[prodmode,acmtecs]{acmsmall} % Aptara syntax
\usepackage{pdfpages} % I ADDED THIS... THIS WAS NOT PART OF THE ORIGNAL TEMPLATE.
\usepackage{rotating}
\usepackage[ruled]{algorithm2e}

\SetAlFnt{\small}
\SetAlCapFnt{\small}
\SetAlCapNameFnt{\small}
\SetAlCapHSkip{0pt}
\IncMargin{-\parindent}

\begin{document}

% Page heads
\markboth{Rafael Roman Otero, Alex Aravind}{MiniOS: an instructional platform for teaching operating systems labs}

% Title portion
\title{MiniOS: an instructional platform for teaching operating systems labs}
\author{
RAFAEL ROMAN OTERO
\affil{University of Northern British Columbia}
ALEX ARAVIND
\affil{University of Northern British Columbia}}
% NOTE! Affiliations placed here should be for the institution where the
%       BULK of the research was done. If the author has gone to a new
%       institution, before publication, the (above) affiliation should NOT be changed.
%       The authors 'current' address may be given in the "Author's addresses:" block (below).
%       So for example, Mr. Abdelzaher, the bulk of the research was done at UIUC, and he is
%       currently affiliated with NASA.

\begin{abstract}
Delivering hands-on practice laboratories for introductory courses on operating systems is a difficult task. One of the main sources of the difficulty is the sheer size and complexity of the operating systems software. Consequently, some of the solutions adopted in the literature to teach operating systems laboratory consider smaller and simpler systems, generally referred to as instructional operating systems. This work continues in the same direction and is threefold. First, it considers a simpler hardware platform. Second, it argues that a minimal operating system is a viable option for delivering laboratories. Third, it presents a laboratory teaching platform, whereby students build a minimal operating system for an embedded hardware platform. The proposed platform is called MiniOS. An important aspect of MiniOS is that it is sufficiently supported with additional technical and pedagogic material. Finally, the effectiveness of the proposed approach to teach operating systems laboratories is illustrated through the experience of using it to deliver laboratory projects in the Operating Systems course at the University of Northern British Columbia. Finally, we discuss experimental research in computing education and considered the qualitative results of this work as part of a larger research endeavour.
\end{abstract}




% We no longer use \terms command
%\terms{Design, Algorithms, Performance}


% At a minimum you need to supply the author names, year and a title.
% IMPORTANT:
% Full first names whenever they are known, surname last, followed by a period.
% In the case of two authors, 'and' is placed between them.
% In the case of three or more authors, the serial comma is used, that is, all author names
% except the last one but including the penultimate author's name are followed by a comma,
% and then 'and' is placed before the final author's name.
% If only first and middle initials are known, then each initial
% is followed by a period and they are separated by a space.
% The remaining information (journal title, volume, article number, date, etc.) is 'auto-generated'.

\maketitle

%%%%%%%%%%%%%%%%%%%%%%%%%%%%%%%%%%%%%%%%%%%%%%%%%%%%%%%%%%
%%%%%     				S E C T I O N			      %%%%
%%%%%%%%%%%%%%%%%%%%%%%%%%%%%%%%%%%%%%%%%%%%%%%%%%%%%%%%%%

\section{Introduction}

Operating Systems is a central topic in undergraduate computer science curricula. Comprehension of subsequent computer science courses relies on proper understanding of operating systems concepts. Whilst this is similar to many other undergraduate courses, what makes operating systems peculiar is the difficulty of delivering laboratory assignments. Due to its complexity and scope, OS courses are delivered in several styles. On one extreme, several universities across the world deliver purely theoretical OS courses. Contrarily, other universities, particularly top western Universities, offer OS courses with a heavy project component.
This is the other extreme, and a more effective way of teaching OS, as it gives students the opportunity for a hands-on experience. (The rest fit in between these two extremes.) In the words of M. Ben-Ari:

\begin{quote}
``Programming is the fundamental activity of computing. As such it must be a major component of courses for students of computing. Courses should not be purely descriptive; instead, they must require students to construct implementations.'' \cite{benari}
\end{quote}

Nonetheless, delivering labs where students write or modify an operating system in a semester is a challenge. Operating systems are large, intricate, concurrent, low-level pieces of software. Writing one requires dealing with: i) asynchronous interrupts; ii) direct access to memory and registers; iii) the inner details of the target computer architecture; iv) the size of the OS itself; and v) the concepts and ideas behind each of the different OS components. Thus, offering the same kind of practical exposure present in some other computer science undergraduate courses is, at best, impractical. 

Several approaches for teaching OS laboratories have been proposed in literature. Given that concurrency and low-level programming (i.e. (i) and (ii) above) are inherent to the hardware platform programming model, efforts in the computer science education community have focused on creating smaller and simpler instructional OS (i.e. they have focused on (iv), (v), and less on (iii)). Continuing in this direction, this work takes the small-size philosophy of instructional OS further, and proposes a minimal system to deliver laboratory assignments. In addition, it attempts to lessen the difficulties that originate from programming a complex machine (i.e. (iii)). Specifically, it does so without opting for either simulated or emulated hardware, nor hiding it behind software abstraction. It instead proposes the use of less complex hardware. Then it combines everything together in a laboratory teaching platform called MiniOS. Lastly, we discuss the effectiveness of the proposed approach and our experience using it over the past few years. 

Section 2 briefly reviews the literature related to this work. It categorizes different approaches and presents where our work stands in relation to it. Section 3 traces the origins of the main difficulties in teaching OS labs in terms of software system, hardware platform, and lack of expertise of students in OS development. Then it uses them as the basis to propose a new solution called MiniOS based on minimal software, minimal hardware, and a guide specifying the construction of the system. Section 4 describes MiniOS and its architecture, discusses the embedded target platform, and elaborates on the accompanying guide. Subsequently, a set of laboratory assignments together with recommendations of its delivery are provided. Section 5 discusses the evaluation of the final product, and the experience in using it to deliver laboratory projects. Lastly, in section 6, we conclude with some remarks and future research directions.

%%%%%%%%%%%%%%%%%%%%%%%%%%%%%%%%%%%%%%%%%%%%%%%%%%%%%%%%%%
%%%%%     				S E C T I O N			      %%%%
%%%%%%%%%%%%%%%%%%%%%%%%%%%%%%%%%%%%%%%%%%%%%%%%%%%%%%%%%%

\section{Related Work}

The teaching of OS lab projects can be broadly classified into four approaches: (i) those where the OS is partially or entirely simulated; (ii) those modifying or extending a full-fledged operating system, either desktop, mobile, or embedded; (iii) those where a toy operating system is built from bare metal; and (iv) those modifying or extending an instructional OS (whether they execute on simulated, emulated, or actual hardware).

Simulation based approaches are attractive as they capture high-level functionality, which can be presented in a visual and intuitive manner. Yet simulations are unrealistic, thereby limiting the learning experience. Conversely, modifying or extending a full-fledged operating system, such as GNU/Linux, does provide the experience of working with a \emph{real system}. This is, however, at the cost of a steep learning curve, which results in students only having time to modify a limited number of components in a superficial manner. It is our opinion that these two methods are inadequate, and we will not consider them further. Our views on production operating systems as teaching tools comply with those found in literature \cite{Survey}. Since our approach relates more with (iii) and (iv), they are described in more detail in the remainder of this section.

\subsection{Building a Toy OS from the ground up}

Building a toy OS from the ground up involves students designing their own simple OS. Out of convenience, a virtual machine (e.g., bochs) is typically used as development platform; though it is possible, with some assistance, to have students execute their OS in actual hardware. Examples of instructional operating systems following this philosophy are the uMPS/Kaya platform \cite{kaya}, the TempOS platform \cite{tempos}, GeekOS \cite{GeekOS}, VIREOS \cite{vireos}, Black's OS \cite{black}, and Chadwick's OS \cite{cambridge}.
The building of a toy OS approach has the following advantages:

\begin{itemize}

\item  There is no pre-existing OS to assimilate;

\item  Building the system from the ground up demonstrates how the system fits together, thereby providing a holistic view of it; 

\item  Letting students build an OS of their own gives them a gratifying feeling (this is from our experience, and similar views are reported in \cite{kaya}).
\end{itemize}
 
The disadvantage, on the other hand, is that students need to work directly with hardware that is intricate and has a steep learning curve. Complex hardware together with the difficulties of writing an OS from the ground up, leaves no opportunity to cover more than a few topics in their most rudimentary forms. Consider the case of Chadwick's OS, where four out of eleven lessons are dedicated solely to controlling the screen. The final OS then not only has little resemblance with a production OS, but is also a tiny toy---in the sense of not being developed enough to have any practical purpose. 

Moreover, the lack of device drivers  availability is a problem for both students and instructors; as 
they constitute a bulk of operating systems code. Writing drivers is a difficult technical task, beyond the skill set of anyone who is not an experienced kernel developer. Without having proper drivers support, it is impossible to go ahead with projects, be they lab assignments or final projects. For instance, in \cite{black}, a lab project on user and kernel mode separation was rendered impossible due to the use of BIOS for accessing I/O. Because of such complexity, some systems such as KayaOS and VIREOS have opted for running on simpler simulated hardware. Although it does bring the complexity down, it is at the cost of realism.  

\subsection{Modifying/Extending an Instructional OS}

In this approach, students are given the task of manipulating an instructional OS; namely, adding functionality or modifying the existing one. Unlike production operating systems, pedagogic ones are more compact. That is, the number of concepts, amount of code, and technical details that must be comprehended are fewer. Examples in this category are: Nachos \cite{nachos}, Pintos \cite{pintos}, PortOS \cite{portos}, BabyOS \cite{babyos}, OS/161 \cite{OS161}, Topsy \cite{topsy}, among others. 
There are some advantages to this approach:

\begin{itemize}

\item  Their smaller size makes them more approachable than production operating systems.
\item  Interaction with hardware is not direct, as interaction occurs with pre-written lower abstraction layers.

\item  They (some more, some less) resemble operating systems as built in reality. 
\end{itemize} 

 On the down side, resembling real operating systems entails complexity. Thus, these systems deal with the issue of how much realism must be traded for simplicity. Consider, for example, the case of modifying a FAT32 file system. Students ought to have some understanding of the file format itself, the actual---often non-trivial---code used to implement it and its interaction with other system modules. On the other hand, a simpler ad-hoc file format, which lends itself to easier comprehension, is not a file format used in deployed systems. In other words, there is no single system that fits both simplicity and realism. Instructors must, therefore, select one that adequate to their teaching objectives.

Like build-your-own approaches, most instructional systems cannot be used for any purpose other than instruction; as often they do not run on actual hardware, or they simply are not developed enough. Those that can are complex systems. A survey of instructional OSes can be found in \cite{Survey}.

\subsection{The Xinu approach}
An intermediate approach that fits in the two previous categories is Xinu \cite{comer}. Xinu is an instructional OS and is peculiar in that a guide (in the form of a book) to its design and implementation is available. While its design and inner workings are detailed, its implementation is also given a rationale and demonstrated in code. This makes Xinu more self-contained; meaning, that the necessary knowledge to put the system together is part of the guide. Altogether, the book removes the mystery surrounding the OS.

Instructors may have options for students to either extend/modify the system or build it in its entirety. So, the advantages and disadvantages are a mixture of the two previous approaches. Importantly, the system developed is not a toy, but a complete and functional operating system, and for the same reason, it is complex. In fact, it is intended to be used for advanced courses with a focus on production operating systems. Consider one of the highlight remarks from the back cover of the Xinu Book (Lynksys version):

\begin{quote}
``Designed for advanced undergraduate or graduate courses, the book prepares students for the increased demand for operating system expertise in industry.'' \cite{comer}
\end{quote}

Similarly, from embedded Xinu's website:

\begin{quote}
``A student built operating system puts the student in the trenches of operating system development. The student will become intimately involved with the inner workings of an operating system.'' \cite{xinuwiki}
\end{quote}

Hence, this approach is less suited for any introductory operating systems course. Further, the guide does not touch on any hardware-related details, leaving to students and/or instructors the task of filling the considerable gap between the OS guide and the hardware documentation. One might point out that other operating systems such as Minix \cite{minix}, Kaya, TempOS and Topsy also come with a document describing the system. Yet, none is as self-contained, nor offers the amount of detail as that of Xinu.

With this brief review of the related work, we next present the rationale for our approach.

%%%%%%%%%%%%%%%%%%%%%%%%%%%%%%%%%%%%%%%%%%%%%%%%%%%%%%%%%%
%%%%%     				S E C T I O N			      %%%%
%%%%%%%%%%%%%%%%%%%%%%%%%%%%%%%%%%%%%%%%%%%%%%%%%%%%%%%%%%

\section{Rationale for MiniOS}

A student-built production-like operating system that can run on real hardware is the ideal under the philosophy of hands-on experiential learning. Similar views are expressed in \cite{kaya} and \cite{OS161}. Whether the system of choice is mobile, desktop, embedded, or some other will depend on the specific instructional objectives of courses. In either case, a dichotomy exists between the ideal and fitting the workload into one semester. In this section, we attempt to explain this dichotomy and 
then provide the basis of our solution.

\subsection{The issues of building a complex system}

Operating systems are complex. Mosley et al \cite{tarpit} point out that complexity has a direct impact on one's attempts to understand a system. They also identify the three main sources of complexity in software: {\sl state, flow of control,} and {\sl code volume}. Given the sizes of modern operating systems, as well as the topic in question (OS instruction), we focus our attention on code volume.

Modern operating system sizes are typically in the order of millions lines of code (LoC), and it is not surprising to find them in the list of the largest softwares \cite{AndroidLOC}. Consider, for instance, the latest versions of the mobile operating systems Android and Symbian, or the latest versions of the desktop systems GNU/Linux, Mac OS, and Windows; all of them are composed of millions of LoC.

For this reason, the computer science education community has favoured the use of smaller instructional operating systems. That is, simpler systems whose main purpose is to serve as a teaching tool. To put this into perspective, consider the size, in LoC, of some popular instructional systems: Minix and Xinu, with tens of thousands; Kaya OS with over 7,000; Pintos with over 5,000; and Nachos with approximately 2,500. Their smaller size enables labs to be carried out in one semester's time (some with more difficulty than others).

Instructional systems may or may not have what we consider two important characteristics. In particular, being complete and functional. Complete in the sense of implementing the typical components of an OS, and functional in the sense of supporting execution of real applications on real hardware (e.g., a teller machine system, a basic laptop, or a robotic system). Unfortunately, a complete and functional system is more realistic, and thereby, more complex. For instance, Minix and Xinu are complete and functional systems; hence, unsuited for undergraduate instruction. From those remaining, none of them are functional, and their degree of completeness varies. Among them, nachos is arguably the smallest yet complete, due to their philosophy of minimal implementations. In fact, we use the term \emph{minimal} to mean exactly what \cite{nachos} describe in their implementation philosophy: ``Our approach was to build the simplest implementation we could think of for each sub-system'' \cite{nachos}.

Accordingly, we argue that, by means of minimal implementations, we can build a system with further reduced code volume. Moreover, we can use this reduction in size and complexity as an opportunity to:

\begin{itemize}

\item[a)] cover (i.e. implement) components that are otherwise ``out of scope'', and 
\item[b)] build a system capable of serving a purpose using actual hardware.
 \end{itemize}
  In other words, we make the case for a minimal---yet complete and functional---instructional operating system.
Thus far we have discussed the difficulties of dealing with OS software. Now, we consider another source of complexity: the target hardware platform.

\subsection{The issues of complex hardware}

Present time computers are intricate pieces of hardware. Manuals detailing the functionings (from a programmer perspective) of a modern 32- or 64-bit processor add up to at least a few thousand pages. To that, one must add the documentation detailing the functioning of the rest of the computer hardware, e.g., interrupt controller, BIOS/UEFI, timer, real-time clock, and others. 
For these reasons, writing non-trivial bare-metal applications (such as operating systems) is a technical, tedious, error-prone, and laborious task. One ought to know the precise inner workings of the computer if one hopes to direct it to do anything. Even though OS courses are customarily preceded by architecture or organization courses, these inner workings are often too advanced, and there are too many details to be covered in their entirety. Additionally, the machine exposes a programming model of asynchronous interrupts. 
Concurrent code accessing arbitrary memory and registers is one of the most challenging code students will encounter during their studies.

Then, how can a student possibly aspire to build an operating system, even a simple one, in one semester? The answer is simple---they cannot. For this particular source of complexity, solutions have been proposed in literature. One solution is to build the system for a hardware simulator or emulator that is simpler to interact with. This is advocated and demonstrated in Kaya OS, OS/161, VIREOS, PortOS, and Nachos. For example:

\begin{quote}
``Simulators are used to eliminate the burden of working on a bare machine, which, given the time frame of a single term, is outside the scope of an undergraduate's ability.'' \cite{kaya}
\end{quote}

A second solution is to abstract away hardware via a software layer. While this is indirectly followed by any instructional OS not meant to be built from bare metal, GeekOS explicitly follows this approach:

\begin{quote}
``Working at the hardware level has two main disadvantages. First, hardware devices can be tricky to program
correctly. A more fundamental problem is that debugging kernel code running on real hardware is difficult, even for experts. The contribution of our work is to show that both of these difficulties can be overcome without requiring heroic measures from students or instructors. We have implemented a tiny OS kernel, called GeekOS, which provides a sufficient abstraction layer over the hardware to hide the genuinely difficult details.'' \cite{GeekOS}
\end{quote}

A third solution is to compromise on the level of sophistication of the system, so as to simplify the technical (hardware) details required to build it. This is put into practice in Black's OS and BabyOS, where students build a toy OS from bare metal. 

We are of the opinion that exposing the students to real hardware is not only essential for a holistic understanding of the system, but also increases their engagement. Similar views are expressed by Pfaff et al \cite{pintos}. Therefore, we consider only the latter approach of compromising on the level of sophistication. Unfortunately, such compromise results in a system that does execute in real hardware, but is far from being complete and/or functional. 

Yet, we argue that by targeting a simpler real hardware platform, we can decrease the technical knowledge required to build an instructional system; then, use that as an opportunity to build a complete and functional system. In other words, we make the case for a minimal---yet complete and functional---instructional operating system for a minimal hardware platform.

Instructional OSes achieve simplicity by trading the capabilities of full-fledged real systems. Next, we elaborate on it.

\subsection{A minimal instructional OS for a minimal platform}

If realism must be traded for simpler minimal implementations, the question that follows is, where is the ideal trade-off point between one and another? This is a difficult question, and it is (directly or indirectly) explored in each and all of the different instructional operating system proposals. For instance, Holland et al elaborate: 

\begin{quote}
``For teaching, a certain amount of realism is desirable. Too much realism, however, becomes both too complicated and, sometimes, realistically painful. [...] [R]eal OSes are immensely large and complicated, and are full of complexities and constructs for coping with real-world issues that have little instructional value.'' \cite{OS161}
\end{quote}

Liu et al also elaborate:

\begin{quote}
``In the process of using BabyOS, we found that it is really difficult to make tradeoff between realism and simplicity. A certain amount of realism is desirable, otherwise BabyOS feels like an unreal OS. Too much realism, however, becomes too complicated and, student would fail to finish their projects.'' \cite{babyos}
\end{quote}

Even though we do not know where the ideal trade-off point resides, it is our intention to explore it by implementing a minimal instructional OS for a minimal hardware platform, which we call MiniOS. 

Real being impractical, we focus on the minutiae that can preserve ``relevant realism'' in trade of ``less relevant realism'' (as far as undergraduate instruction goes). In particular, MiniOS is complete and functional; it is targeted for a real hardware platform; and it follows the design, layout, and mechanisms of real systems. Meanwhile, fault tolerance, robustness, efficiency, reliability, sophistication, and other attributes in deployed systems are not considered. 

To put it bluntly, whilst MiniOS should not be deployed as part of an aircraft computer or an X-ray device, it is perfectly suitable for less important applications, such as a gardening system, or a small mobile robot---and such system, we believe, is well suited for instruction.

Thus far we have used the term \emph{real hardware} generically, now it is time to specify a target platform. 

\subsubsection{A minimal embedded hardware platform}

We use the term minimal hardware to refer to those computers with the least amount of sophistication still capable of hosting an OS. For the sake of exploration, we have selected what we consider to be one of the smallest among them; more specifically, a 32-bit ARM low-end embedded platform. This choice is partly arbitrary and partly influenced by ARM's popularity in the mobile and embedded systems industries.

It is worth noting that the term \emph{embedded} does not imply simplicity. While there exists basic 8-bit microcontroller (MCU)-based embedded computers (e.g. a coffee maker's computer), there too exists sophisticated 64-bit microprocessor (MPU)-based embedded computers (e.g. an industrial robot's computer). 

Despite the fact that some instructional systems, such as Minix, Xinu, and BabyOS are targeted for (or have been ported to) embedded platforms, they differ from our philosophy of minimal hardware. In fact, to our knowledge, there is not an existing instructional OS with similar views on hardware.

It is also important to clarify that MiniOS is not intended to be an embedded production OS. Like desktop systems, embedded production OSes are complex. They tend to be plagued with intricacies that make them adequate for deployment in life-critical applications such as aircraft and military. A representative sample, and in the smaller side of the size espectrum, is FreeRTOS \cite{freertos}; which, intended for low-end embedded platforms, has over 9,000 LoC \cite{FreeRTOSLOC}. 

A low-end embedded OS may seem as an over simplification, and naturally one raises the question of whether such a simple system has any instructional value outside the embedded systems realm. 

\subsubsection{A low-end embedded OS as a teaching tool}

The purpose of MiniOS is not to serve as a tool for teaching embedded systems, but to serve as a tool for teaching general principles that apply to operating systems. In fact, MiniOS is not well suited for teaching labs in embedded systems, as embedded-specific details are deliberately overlooked. With few exceptions where it is impossible, it is emphasized how they contrast with general purpose computers. Consider, for instance, the case of a MCU-based low-end embedded platform (a Hardvard architecture) with Flash as program memory; it must be brought to the students' attention that general purpose systems do not, customarily, have non-volatile program memory in their address space. Thus a boot-loader for a MCU will be different than one for, say, a desktop computer.

Fortunately, the similarities are greater than the differences, and this is why we believe a simpler low-end embedded system can be used as a teaching tool. That is, for a course with no intention of preparing students for real-world OS development (whether embedded, desktop, or other).

One important benefit of working with MCU-based embedded platforms is the availability of device drivers. Hardware manufacturers typically release open source bare-metal middle-ware (mostly drivers) to be used on their platforms.

Finally, we argue that, recently, there has been a switch from traditional desktop systems to mobile and embedded systems (e.g. internet of things and wireless sensor networks). An embedded instructional system with wireless capabilities can be a tool for introducing students to the latter. A similar argument is expressed by Atkin and Sirer \cite{portos}.

An equally important aspect of MiniOS is its guide. It covers building the system from nothing, and it is described in the following section.

\subsection{From the ground-up: a guide to MiniOS design}

MiniOS is intended to be built from the ground up, on bare metal. For this, a guide to its design is primary. In a comprehensive and thorough manner the guide must---step by step---detail the construction of the system from nothing. All the technical details dealing with the hardware, the compiler, as well as OS concepts and their specific implementations should be covered; including details such as exceptions, memory mapped IO, linking of relocatable code, calling conventions, memory segmentation, and context switching. 

Other instructional systems also advocate for the use of a guide or manual \cite{GeekOS}, \cite{topsy}, \cite{kaya}, \cite{tempos}, \cite{vireos}, \cite{minix}, \cite{pintos}, and \cite{nachos}; some with more details and code than others. None, however, go to the amount of detail (instruction) that we consider necessary for building an OS from the ground up. (XINU is the exception; the amount of instruction offered as written material in \cite{comer} is near to what we advocate for.) 

Guzdials \cite{guzdial} argues that the amount of instruction matters when teaching computer science to beginners. In particular, ``putting introductory students in the position of discovery information for themselves is a bad idea.'' Although this argument is given in the context of introductory programming (100 level courses), the instruction in question (operating systems) is introductory, as both systems programming and programming at such low-level of abstraction are substantially different from what students have encountered in preceding courses. 

From experience we have noticed that, at this introductory stage, most students lack the experience, the patience, and the right approach to meticulously construct and debug low-level systems' code. Moreover, they are faced with programming patterns and tricks specific to the machine's programming model. While many of these patterns are simple and of common use, it can be difficult to re-invent them if one has never encountered them before. Also, in contrast with higher-level programming, bugs manifest differently (typically the CPU faults and does nothing) in low level. Code is highly dependent on a great number of machine-specific details, all of which must be set correctly, and access to raw memory requires precise knowledge of its organization and how instructions access it. Moreover, it is practically impossible for students to obtain all of the required details for OS construction from the thousands of pages included in the documentation, for they are not at the level of understanding the technicalities therein. The end result is that students are prone to get hopelessly stuck. 

Consequently, we consider that a guide demonstrating how to build the system from the ground-up: as well as specifying, in a comprehensive manner, the technical details relevant to OS writing is a necessity for the delivery of OS labs. With this background, we now proceed to describe our proposed system called MiniOS.

%%%%%%%%%%%%%%%%%%%%%%%%%%%%%%%%%%%%%%%%%%%%%%%%%%%%%%%%%%
%%%%%     				S E C T I O N			      %%%%
%%%%%%%%%%%%%%%%%%%%%%%%%%%%%%%%%%%%%%%%%%%%%%%%%%%%%%%%%%

\section{MiniOS---Proposed OS instructional platform}

The proposed OS instructional platform consists of the system, the target hardware, and its construction guide. This chapter describes them and gives a set of suggested laboratory projects, as well as recommendations for their delivery. 

\subsection{The system}

First we present the high level architecture of the system, and then describe the different parts that constitute the system.

\subsubsection{Architecture}

\begin{figure}
\centerline{\includegraphics[scale=0.45]{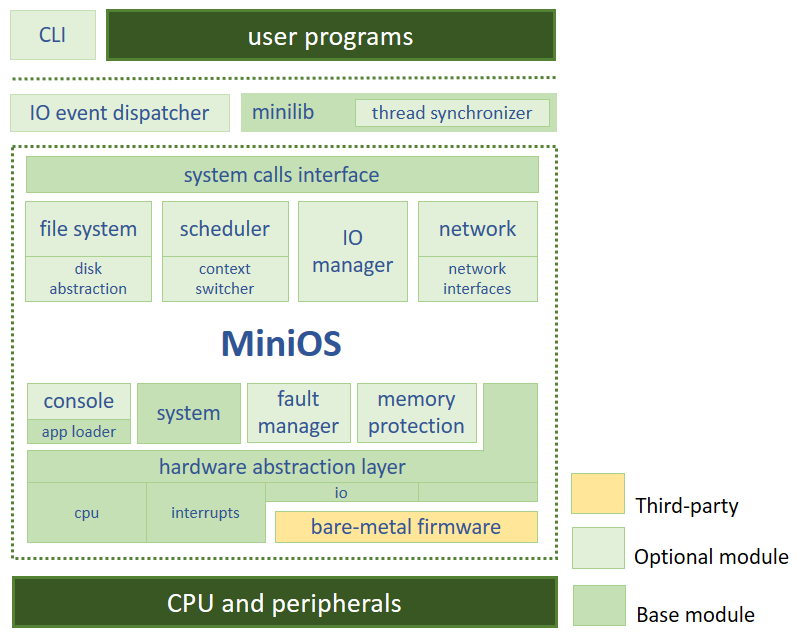}}
\caption{MiniOS' Architecture.}
\label{fig:arch}
\end{figure}

From an architectural point of view it is unclear the parts that must be included in a presumably minimal, complete, and functional operating system. It cannot be composed of too many parts (layers or modules) as to become too complex, nor it should have too few as to be incomplete or non-functional. Our approach on this is to incorporate components typically found in production systems, and offer the choice of what components make it into the system. Specifically, the system is built as a set of loosely coupled modules categorized in base modules and optional modules, as shown in Figure~\ref{fig:arch}. 

As their names suggest, \emph{base modules} form the foundation of the system and must be implemented, whereas \emph{optional modules} add specific functionality that may or may not be integrated in the system. This configuration gives lab instructors and students the flexibility to start with a minimal base and add modules to accommodate to their instructional objectives. 
Complying with the minimality principle, the system has as few lines of code and as few components as possible. 

From a design perspective, we classify the modules into two types: {\it primary} and {\it secondary}.  \emph{Primary modules} represent an identifiable OS component: hardware abstraction layer (HAL), fault manager, memory protection, file system, scheduler, IO manager, network stack, system calls interface, IO event dispatcher, minilib, thread synchronizer, and command-line interface (CLI). \emph{Secondary components} offer some abstraction or functionality but do not represent an OS components: context switcher, disk abstraction, network interfaces, app loader, IO, CPU, and interrupts. Every component is mapped to a source file of the same name. There are as many C or assembly files as there are components in the architecture. 

From a software engineering perspective, a modular architecture has additional benefits. First, it improves modifiability of the system. It allows students to add or remove modules with little or no modification of others. Second, it improves local reasoning, hence aiding our main objective of making the system easier to comprehend. Such design is typically achieved with support from a programming language. However, since the system is written in C and assembly, we rely merely on discipline. Particularly, we strongly advise students to keep state confined to the scope of a module, and let module interaction occur only via interfaces; practices, which we demonstrate throughout the construction guide. It is worth noting that, although instructional systems are more or less designed in this manner, often modules end up keeping global state used by other modules. More importantly, we want to make it explicit that these software engineering practices are essential for our purpose.

An important aspect of MiniOS architecture is that, unlike production systems, device drivers are in direct contact with hardware. This means, the system is built on top of them, instead of them being part of the system itself. Although some instructional systems follow this design for simplicity purposes, we do it explicitly to support integration of open source third-party firmware that is often only available as bare-metal. With this small design choice, MiniOS benefits from available code from chip vendors and/or embedded systems enthusiasts. With this higher level description, next we will describe the individual components.

\subsubsection{Components description}

Let us begin with base modules.

\begin{itemize}

\item {\sl HAL.} This is the lowest layer of the system and it is responsible for providing sensible machine-independent abstractions to upper layers. Particularly, it implements three abstractions: CPU, interrupts, and IO. 

\item {\sl System.} System is central to the rest of the modules, and is in control of all the system-related tasks, such as system initialization and kernel panics. Additionally, it offers implementations of various data structures to aid in the development of the kernel. 

\item {\sl Application loader.} This module is responsible for the loading of applications from the SD Card. 

\item {\sl Console.} The Console is a kernel shell that runs after system initialization. It allows to run applications by name. In addition, it supports a small number of commands, such as \emph{ls}, \emph{cd}, \emph{cat}, and \emph{netstat}; minimal versions of GNU/Linux's commands with the same name. 

\item {\sl System calls interface.} After configuring the CPU to run in user mode, the system calls interface serves as the only gateway to the system. Invocation of system calls is via software interrupts. 

\item {\sl Minilib.} This small library module sits in between applications and system calls. Minilib's purpose is to wrap up system calls and present user applications with a more intelligible interface. It also provides support for buffered IO operations in the presence of the IO manager.

\end{itemize}

Now we describe optional modules.

\begin{itemize}

\item {\sl Fault manager.} It is a small module whose only task is to raise kernel panics on the occurrence of CPU faults (e.g. div-by-zero fault).

\item {\sl Memory protection module.} This module protects kernel code and data from code running in user mode. It restricts applications from accessing specific parts of memory, generating a segmentation fault if boundaries are violated. 

\item {\sl Thread synchronizer.} Albeit part of minilib, thread synchronizer is a module on its own. It contains common implementations of thread synchronization mechanisms: lock, semaphore, monitor, and barrier synchronizations. 

\item {\sl Scheduler.} The scheduler is a limited, but functional, priority-based pre-emptive thread scheduler. It supports a fixed number of threads with fixed stack sizes. While termination for a given thread is supported, freeing of its memory is not (mainly to avoid handling complex memory details). Threads can yield, can signal other threads, and can sleep. For portability, platform-dependent code for context switch is part a context-switcher, and not the scheduler itself.

\item {\sl File system.} A functional operating system must have a file system to start with. MiniOS uses FatFS \cite{fatfs} as file system. FatFS is a small FAT file system for resource-constrained devices.

\item {\sl CLI.} The command-line interface is a user-level shell whereby users can access kernel services. 

\item {\sl Network.} As for networking capabilities, the network stack supports a very simple, inefficient, but functional network protocol over IEEE 802.15.4. Namely, it uses a flooding algorithm (Trickle \cite{trickle}) to form a network of ad-hoc connected devices. To avoid dependencies, the network stack purposefully overpasses the IO manager and handles its own buffers and radio interrupts.

\item {\sl IO manager.} The IO manager controls access of I/O devices. When interrupt-based devices notify the system of available data, it is responsible for a) placing the incoming data in an intermediate buffer accessible to both minilib and the IO event dispatcher; and b) notifying the scheduler of new incoming IO data. 

\item {\sl IO event dispatcher.} The IO event dispatcher enhances the system with \emph{IO events}. Whenever the scheduler is notified of new incoming IO data, the event dispatcher runs and executes the corresponding user-level event handler. Unlike other modules that can be implemented on top of base modules, the IO event dispatcher requires the scheduler and the IO manager to be part of the system.

\end{itemize}

For a more concrete idea, consider the following sample program.

\begin{verbatim}
#include "minilib/thread.h"
#include "minilib/display.h"
#include "minilib/network.h"
#include "minilib/sensor.h"
#include "minilib/led.h"
#include "minilib/ioevents.h"

void salute_thread( void* params ){
   thread_set_priority( (uint32_t)params );
	
   while( true ){
   	  //print salute to USB
      usb_write( "Hello, I'm thread %s \n", thread_get_current() );
      thread_sleep( 200 );
   }
}

IOEvent io_in_network( NetFrame* frame ){
    //echo
    net_write( frame );		
}

int main(){
    //Create salute threads
    thread_create( salute_thread, "Mariana", 128, THREAD_PRTY_MIN );
    thread_create( salute_thread, "Cafe", 128, THREAD_PRTY_MIN );	
	
    uint32_t state = 1;	
	
    while( true ){
        //print sensor information to OLED display
        display_write( "Light level (%%): %d \n", sensor_light_read() );
        display_write( "Temperature (C): %d \n", sensor_temp_read() );

        //blink LED0		
        led_write( Led0, state++ % 2 == 0 ? LedOn : LedOff );
		
        thread_sleep( 500 );
    }
}
\end{verbatim}

This program is composed of four threads, one of which is main. Two of them print their name approximately five times a second over a CDC USB connection; one waits for an incoming network message and echoes it back to the same sender; and main prints sensor information on the OLED screen and blinks an LED approximately every half second.
 
An operating system works closely with a specific hardware. The following section discusses the target hardware platform.

\subsection{The target hardware platform}

Among all the different available ARM processor cores on the market, the Cortex-M series are those with the least sophistications that still offer support for operating systems. Among them, we have opted for the Cortex-M4, which was the most sophisticated in the Cortex-M series at the time MiniOS was initially conceived. Some of these OS-supporting features are software interrupts, memory protection, different CPU modes (kernel and user), separate user and kernel stacks. In fact, the only missing feature to fully support a conventional OS, capable of executing applications, is a memory management unit (MMU).

\begin{figure}
\centerline{\includegraphics[scale=.35]{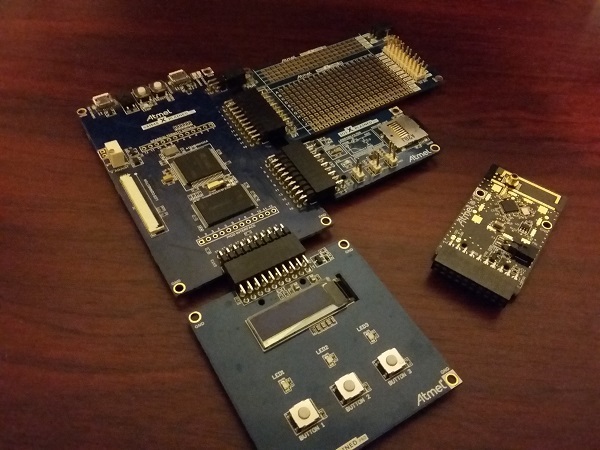}}
\caption{MiniOS' target hardware platform.}
\label{fig:platform}
\end{figure}

Cortex-M cores are only available in MCUs, and because a MCU by itself is of no use, a MCU prototyping (evaluation) board must be used. Although it is possible to carry out labs with tailor-made hardware, an off-the-shelf board has its advantages. First, there are available device drivers from manufacturers. Second, these boards typically integrate an on-board chip debugger and programmer, thereby eliminating the need of an expensive JTAG emulator that does the same. Third, they can be purchased by anyone interested in taking or delivering the course. Lastly, being official boards, they integrate seamlessly with manufacturers' development tools.

A variety of MCU prototyping boards exist in the market from different vendors. Based partly on its low cost, and partly in nothing in particular (as they all are quite similar), we have selected the \emph{Atmel SAM4S Xplained Pro Starter Kit}. Its main board runs at 120 Mhz, and together with its three expansion boards integrate enough peripherals for laboratory projects. They include a small OLED screen, buttons, LEDs, a light sensor, a temperature sensor, a microSD card slot (and the microSD card), a USB device port, an on-board 256 MB Flash memory, and exposed pins for various on-chip peripherals. Figure~\ref{fig:platform} shows the main SAM4S board and its daughter boards, together with the \emph{REB233 board} (acquired separately) for IEEE 802.15.4 connectivity. This is the hardware assumed by the construction guide. 

Being an embedded platform, development of software is somewhat distinct. The following section attempts to offer more details in this regard.

\subsubsection{Development Environment}

\begin{figure}
\centerline{\includegraphics[scale=.35]{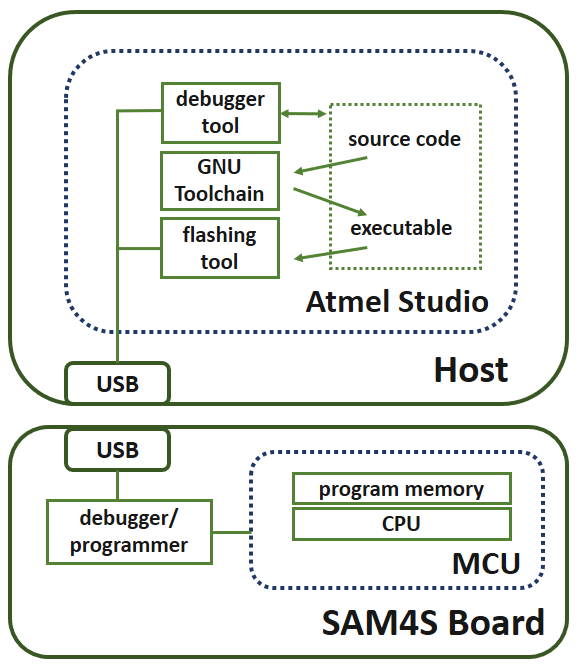}}
\caption{MiniOS development environment}
\label{fig:devenvironment}
\end{figure}

Clearly one cannot (easily) use the system's target platform to develop the system itself. Instead a separate \emph{host computer} is necessary for development of the system. In particular, using a cross-compiler, first the source code is compiled to an executable in the host. Then, the executable is flashed to the target's program memory by a flashing tool. Finally, for debugging, an on-board hardware debugger interfaces with software in the host to enable source-level debugging. All of the different host-side software tools, including the GNU toolchain are integrated in Atmel's IDE: Atmel Studio. Communication between the target platform and host tools is via USB. Figure~\ref{fig:devenvironment} depicts the described programming environment. Incidentally, Atmel Studio was built with Microsoft Visual Studio Shell. So, the programming environment is the same as that from Microsoft Visual Studio.

The Atmel debugging facilities, when used correctly, allow debugging of firmware running in the MCU as if it was a regular desktop application. It allows pausing (possibly at breakpoints) of the CPU for inspection and modification of memory, registers, IO interfaces, and source-level variables (including not primitive types). It also allows to step through both assembly and C code, as well as dis-assembled code. 

The final piece in the development platform is the system's guide to its construction, which is discussed in the following section.

\subsection{The MiniOS Book}

The idea of the MiniOS guide (or book) is to:
\begin{itemize}
 \item Cover in-detail all the technical material that is necessary to build MiniOS.
 \item Guide students in the process of developing it themselves. 
\end{itemize} 
Consequently, the guide is intended to be self-contained, in the sense that a student could rely solely on it to build the system (characteristic not present in other similar OS books).
  The style of the guide was initially inspired by the tutorial \emph{Write Yourself a Scheme in 48hrs} \cite{scheme}, and later by the more textook-like style of the Xinu Book \cite{comer}.
Our guide is divided in two parts, as shown in Table~\ref{tab:book}.

The intention of the first part is to instruct on computer architecture using the ARM Cortex-M4 and the SAM4S board. The second part is dedicated entirely to the system, and it assumes some working knowledge of what is covered in the first section. Ideally, a student should complete the first section of the book, and then engage in building the system. However, if this is not the case (as we have experienced), working knowledge of a different computer architecture suffices. At worst, students will take extra time to learn certain Cortex-M4 technicalities. Importantly, all these required technicalities are available for consultation in the architecture section, and when used in the systems section, they are referenced.

\begin{table}%
\tbl{Book Layout\label{tab:book}}{%
\begin{tabular}{|l|l|}
\hline
\multicolumn{1}{|c|}{\textbf{PART I}}       & \multicolumn{1}{c|}{\textbf{PART II}} \\
\multicolumn{1}{|c|}{\textbf{(HW ARCHITECTURE)}} & \multicolumn{1}{c|}{\textbf{(SW SYSTEM)}}  \\ \hline
1. Introduction        					  & 1. Basic IO and Booting      		 \\
2. Instruction Set Architecture           & 2. Hardware Abstraction Layer        \\
3. Memory                                 & 3. Executing Applications            \\
4. IO									  & 4. System Calls                    \\
5. Stack                           		  & 5. Fault Manager                    \\
6. Interrupts                             & 6. Memory Protection               \\
			                              & 7. Scheduler                 		 \\
                                          & 8. IO events        				\\
                                          & 9. Thread Synchronization 		 			\\
                                          & 10. Network Stack           		\\ 
                                          & 11. Command-Line Interface           \\ \hline                                                  
\end{tabular}}
\end{table}%

An important aspect of the guide is the great amount of details offered. This is because it was written with the purpose to not leave students in the situation of discovering neither advanced topics nor topics pertaining to other subjects by themselves. Among others, it covers topics and information related to data structures, drivers, CPU, peripherals, the SAM4S board, the SAM4S MCU, the C language, assembly, the linker, and even programming patterns that are particular of systems or low-level programming. For instance, the guide explains and demonstrates the following: how to use callbacks to push data (coming from interrupts) from a lower layer to an upper layer; how to load a pre-compiled application from permanent storage to RAM for execution; how to write a linker script; how to do context switch; how to change CPU privileges; where in the documentation to find the mapping between physical pins and logical IO bits; and so forth. Some of this information is too technical or advanced to be left for discovery, and some does not pertain to OS instruction per se. Appendices ~\ref{sec:appendix-apps} and ~\ref{sec:appendix-syscalls} show excerpts from the Executing Applications and System Calls Chapters. 

Importantly, most pieces of code that are given, are not just given, but derived, meaning, the book explains the steps in obtaining it from documentation or other assumed background knowledge. This gives interested and motivated students the tools to modify those parts, should they want to (e.g. for a final project).

Additionally, a device driver integration section was added to the guide. This sub-guide demonstrates the process of integrating third-party drivers, and shows working sample code. It can be challenging to write working code for an IO peripheral out of poor, and often buggy or incomplete, third-party documentation. Appendix~\ref{sec:appendix-drivers} shows one sample entry from the device drivers section.

\begin{figure}
\centerline{\includegraphics[scale=.3]{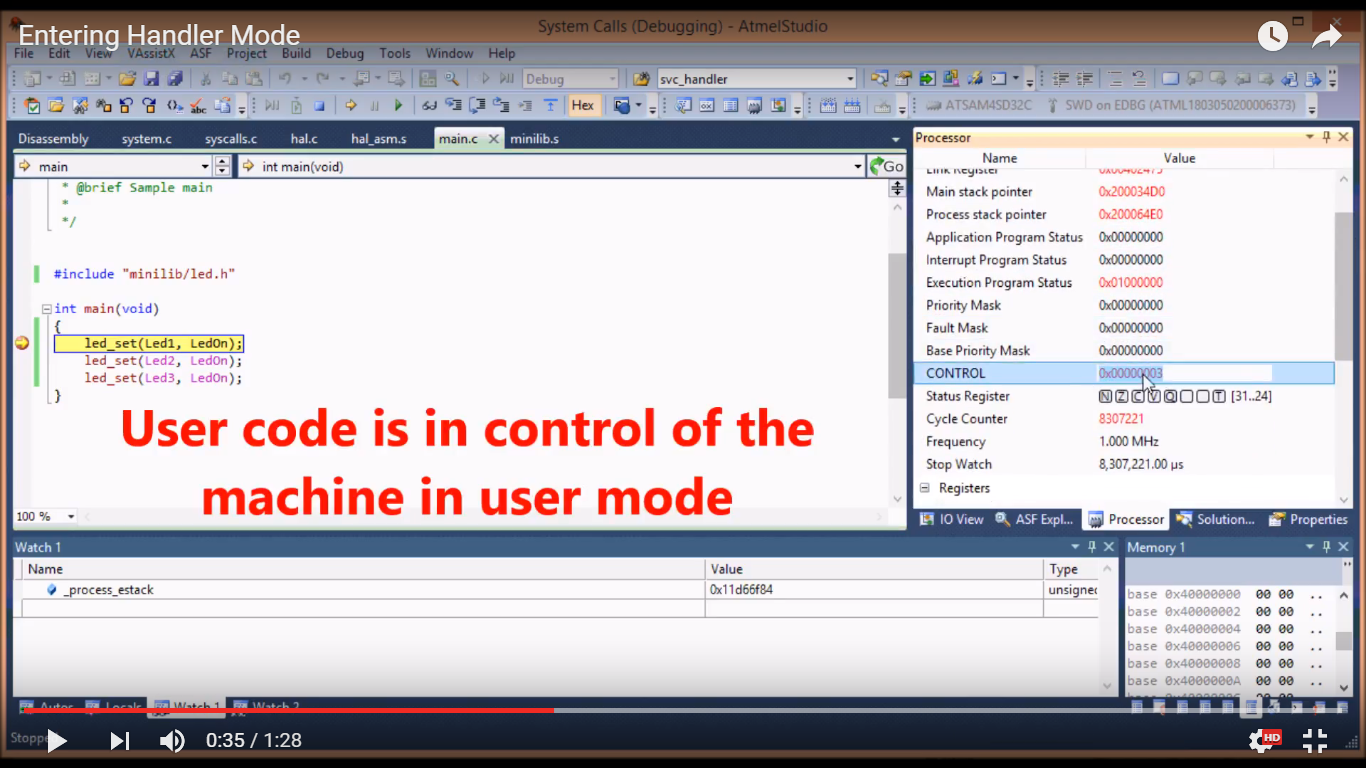}}
\caption{Video demo: Entering kernel mode}
\label{fig:samplevideo}
\end{figure}

In addition to text material we have prepared demonstrative videos. These are videos made to strengthen the text material, by showing explained concepts, techniques, processes, solutions to labs, or running sample driver code. For example, Figure~\ref{fig:samplevideo} shows a debugging session right before a system call. In this video, the control register is highlighted to demonstrate that, in fact, the CPU is in both user and unprivileged mode. Upon execution of the software interrupt, it is shown again, but this time specifying kernel privileged mode. 

Lastly, we would like to emphasize that all this extra instruction goes in accordance with what is argued by Guzdial et al in \cite{guzdial} in favour of strong instructional guidance for novice learners. In particular, he mentions that there is  strong evidence that the minimal guidance approach we typically use in computer science instruction is inadequate. One can argue that operating systems and computer architecture are typically second and third year courses, and therefore students are not novice programmers. While this is true, students are still considered novice learners from a low-level programming and OS development perspective.

Based on the described system, hardware platform, and guide thus far, the next section presents suggestions on how to accommodate the material in actual laboratory projects. 

\subsection{Laboratory Projects}

There is a total of twelve labs, with different suggested durations. The first lab is an optional short introduction. The next two labs are also short, and, since they involve base modules, they cannot be skipped nor their order can be altered. The remaining eight are optional, and most of them can be implemented regardless of order. In case a module is considered to be good to have, but not of interest as to dedicate a lab to it, there is the possibility to hand it in to students. For instance, the fault manager can be a useful module to have as it outputs human-readable messages when the CPU faults, and it could be given to students.

\subsubsection{Lab 1 - Basic IO and Booting}

\begin{figure}[ht]
\centerline{\includegraphics[scale=.3]{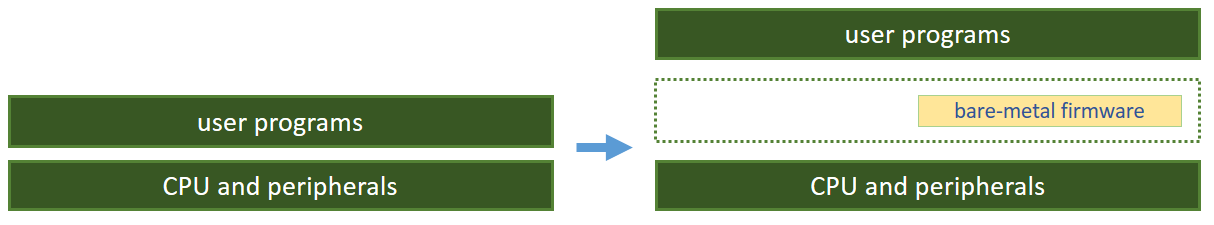}}
\caption{Architecture Goal for Intro Lab}
\label{fig:lab1}
\end{figure}

In this lab students are introduced to the booting process, the use of third-party firmware as basic input-output, and the programming environment (including debugging facilities). The recommended time for solving this lab is one week and is optional, albeit recommended. 
Another way of looking at this lab is that it enhances bare-metal applications with bare-metal firmware, as depicted in Figure~\ref{fig:lab1}.

The learning outcomes for this lab are to familiarize students with the development environment; to provided some guidelines on how to make efficient use of the debugging facilities; and to show the process of integrating third-party firmware to be used as basic IO.

\subsection{Lab 2 - Hardware Abstraction Layer}

\begin{figure}[ht]
\centerline{\includegraphics[scale=.3]{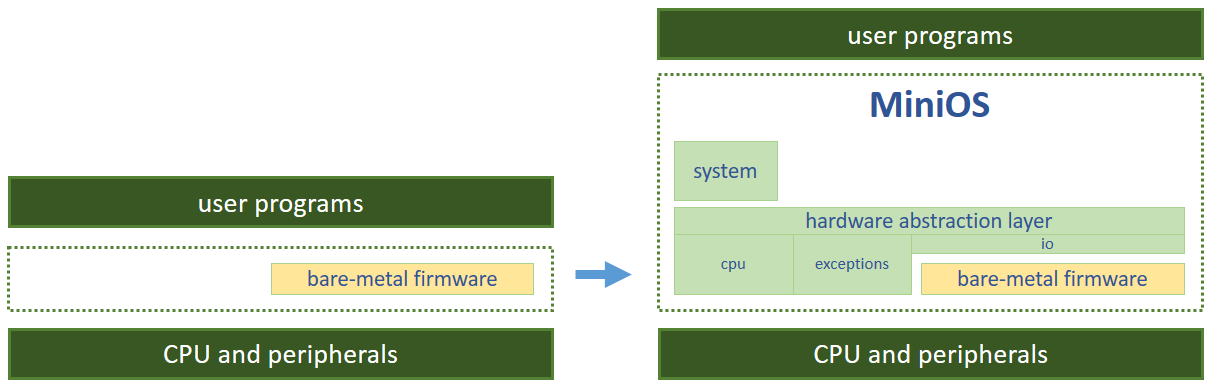}}
\caption{Architecture Goal for HAL Lab}
\label{fig:lab2}
\end{figure}

For this lab students write the HAL, and the system module. Some of the implementations expected from this lab are, for example, an IO device abstraction composed of a read function and a write function, an abstraction for registering callbacks of interrupt-based IO, among others. The recommended time for solving this lab is one week, and it is mandatory. Figure~\ref{fig:lab2} shows the result of completing the HAL.

The objective of this lab is to give some insight and hands-on experience on interaction with bare-metal IO peripherals, and in the process convey students the importance that a) abstraction plays in development of the system, and b) the repercussions that a HAL has in portability. 

\subsection{Lab 3 - System Calls}

Provided hardware-specific information on how to establish a kernel and user mode separation, students must add code to support software-interrupt-based system calls via minilib. The separation is made even clearer by splitting compilation of OS and app. MiniOS is compiled and flashed to the MCU, while applications are compiled and moved to an SD Card from where they are loaded into RAM and executed. Since loader code is given, students are asked to write a rudimentary version of the console that supports listing of files and execution of applications. The console is a serial terminal, thus can be accessed via any serial terminal (e.g. PuTTY). Special emphasis is placed on conveying the importance of CPU mode separation in security. In fact the chapter begins by demonstrating an app that changes the user name displayed int he console by directly manipulating a kernel data structure. 

Optional tasks involve writing of various different system calls (e.g. execv for child process creation), or taks related to buffered output (e.g. implementation of a line-buffered display\_write function). In the process,  the inability of user code to directly access data from interrupt-based input is emphasized; although nothing is done about it until later labs.
The recommended time for solving this lab is one to two weeks.

\begin{figure}[ht]
\centerline{\includegraphics[scale=.3]{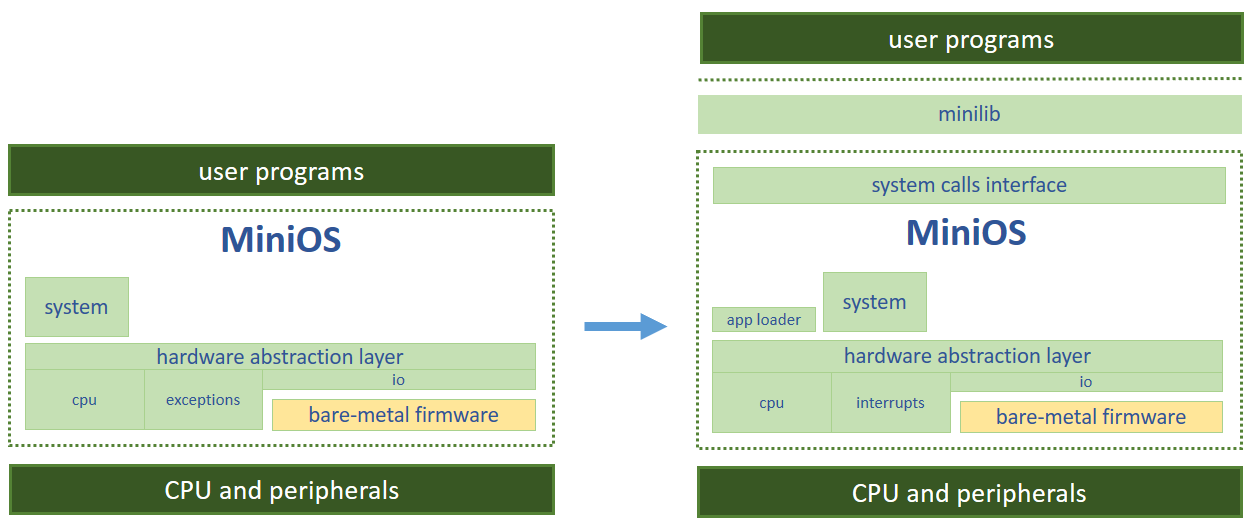}}
\caption{Architecture Goal for System Calls Lab}
\label{fig:lab3}
\end{figure}

Upon completion of this lab students are expected to have some insight and working knowledge of:

\begin{itemize}
\item The need and implementation of kernel-application separation, as well as the mechanisms used by operating systems to interface them both.

\item The role and place of libraries such as the GNU C Library in an operating system.
\item  The limitations of poll-based IO.
\item  Buffered IO.
\end{itemize} 

The resulting architecture is in Figure~\ref{fig:lab3}.

\subsubsection{Optional Lab -- Fault Manager}

The fault manager is another short lab. Here students are given guidance on CPU faults, and are asked to add support for fatal system errors---the mini black screen of death. If memory protection is part of the MiniOS version for this lab, a more complex task involves termination of the offending application and continue of execution. The recommended time for solving this lab is one week. 

For this lab, students are expected to gain insight on error detection, and have some experience in the process of handling and reporting them.

\subsubsection{Optional Lab -- Memory Protection}

For this lab students must implement memory protection to prevent user code from accessing system code and data in memory. If the fault manager has been implemented, an extra task of enabling segmentation faults is available. The recommended time for solving this lab is one week.

The idea of this lab is to supply students with insight and working knowledge on the use of memory protection to prevent bugs and malicious code to mess with the system, as well as let them experience first hand what a segfault is. 

\subsubsection{Optional Lab -- Scheduler}

The scheduler is perhaps the most technically challenging lab. Starting from the system timer, a single-threaded scheduler and a yield function are derived and demonstrated. Available tasks include extending it to support multiple threads, priorities, round-robin scheduling, a sleep function, thread signalling, and different scheduling policies, among other tasks. The recommended time for solving this lab is two weeks, or three if extra tasks are added.

The learning outcomes for this lab are to provide students with experience of the obscure inner workings of a thread scheduler, to let them experience first hand how sharing CPU is made possible by a set of small clever tricks done by the operating system; also, to get some working knowledge on a) implementation of different scheduling policies; b) how different threads queues can be used to support sleeping threads, priorities, and thread signalling, among other tasks.

\subsubsection{Optional Lab -- IO events}

In this lab students write the IO manager and the IO event dispatcher to add support for user-level run-to-completion IO events. Every time new data is received from interrupt-based IO, the IO manager stores it in intermediate kernel buffers and notifies the scheduler to wake up and run the event dispatcher thread. This is a short lab, and its recommended time is one week.

On completion of this lab, students are expected to have an understanding of a) the implementation of events from threads; b) the benefits of a hybrid thread-event approach to concurrent programming.

\subsubsection{Optional Lab -- Thread synchronization}

Here students implement thread synchronization mechanisms: locks, semaphores, monitors, and barrier synchronizations. The recommended time for solving this lab is one week or two depending on the number of mechanisms implemented and the amount of help offered.

At the end of this lab students are expected to understand, from an implementation perspective, synchronization mechanisms used in multi-threaded programming.

\subsubsection{Optional Lab -- Network Stack}

A MAC layer interface is delivered as part of this lab's material. Students must, then, use the trickle algorithm \cite{trickle} to enhance the system with network capabilities. A more complicated task includes writing a host application that transmits an executable, having a node receive it and execute it.  
Since trickle is straightforward to implement, the recommended time for solving this lab is one week, or two with more complicated tasks.

The objective of this lab is to demonstrate, in a rudimentary manner, how computer networks are built out of layer-2 point-to-point communication; also, to show the difficulties of a) providing applications with networking services, and b) dealing with unreliable wireless communications.

\subsubsection{Optional Lab -- Console and CLI}

In this lab students write either a more advanced console (than what they have in Lab 3) or a CLI, or both. The console is internal to MiniOS, it is launched on system start up, and enables:

\begin{itemize}
\item to print information during system initialization; 
\item user login;
\item execution of basic commands; and
\item to browse and execute applications stored in the SD Card.
\end{itemize}
  
System initialization messages include CPU speed and peripherals found. More advanced features involve basic managing of user accounts, and enabling applications to exit and give control back to the console. The CLI, on the other hand, is an application that runs in user mode and allows similar functionality. More advanced tasks include the implementation of privileges for user accounts; replacement of PuTTY for a custom made terminal, so as to give the terminal a more traditional feeling; or a host panel board that shows sensor information.

This is a short lab and the recommended time for solving this lab is one week. The purpose of this lab is to demonstrate how command-line interpreters, whether they run in kernel or user mode, fit in with the rest of the system. 

\subsubsection{Final Project}

As final projects, students may form teams and create something of their own. Any idea involving an embedded OS, or extension of the OS itself are eligible choices. Unlike all the remaining labs, this project has no rigid specification. It is open ended and students are encouraged to implement something of their interest. A complete version of the system can be handed in to those teams who need it. In the end, exact specifications differ depending on instructors' preference. 
The idea is for students to put all acquired knowledge to practice and hopefully deepen their knowledge in some specific OS aspect of their choice. 

This concludes the presentation of the instructional platform. 

%%%%%%%%%%%%%%%%%%%%%%%%%%%%%%%%%%%%%%%%%%%%%%%%%%%%%%%%%%
%%%%%     				S E C T I O N			      %%%%
%%%%%%%%%%%%%%%%%%%%%%%%%%%%%%%%%%%%%%%%%%%%%%%%%%%%%%%%%%
\section{Evaluation}

The idea of MiniOS was not to replace other instructional systems, but to create an alternative system that could adhere itself to the already existing set. More precisely, it was meant to be a small, complete, and functional MCU-based system that could be used for teaching operating systems concepts, while lessening students' struggle.

Due to our policy of minimal implementations and minimal hardware, we were able to write the entire system in less than 3,500 lines of C code and less than 250 lines of assembly. Following the principle of not placing students in the position of discovering new information, we have developed a self-contained book covering the construction of the system. All this together has enabled MiniOS to be:
\begin{itemize}
\item functional and complete
\item small
\item built-from the ground up, and
\item simple enough to reduce students' struggle in building an instructional system. 
\end{itemize} 

For further evaluation, next we report our observations in using MiniOS to deliver operating systems labs at our university. In addition, we present student's feedback, sample final laboratory projects (showing their quality), and student's laboratory grades. Finally, we elaborate on the use of MiniOS as a prototyping and research platform.

\subsection{Observations}

\begin{figure}[ht]
\begin{minipage}{.33\textwidth}
\centerline{\includegraphics[scale=.75]{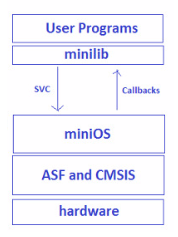}}
\subcaption{(a) 2013}
\end{minipage}%
\begin{minipage}{.33\textwidth}
\centerline{\includegraphics[scale=.28]{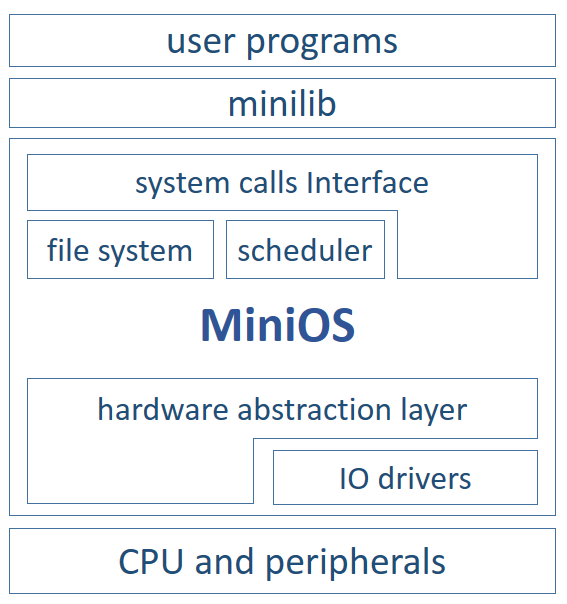}}
\subcaption{(b) 2014}
\end{minipage}%
\begin{minipage}{.33\textwidth}
\centerline{\includegraphics[scale=.28]{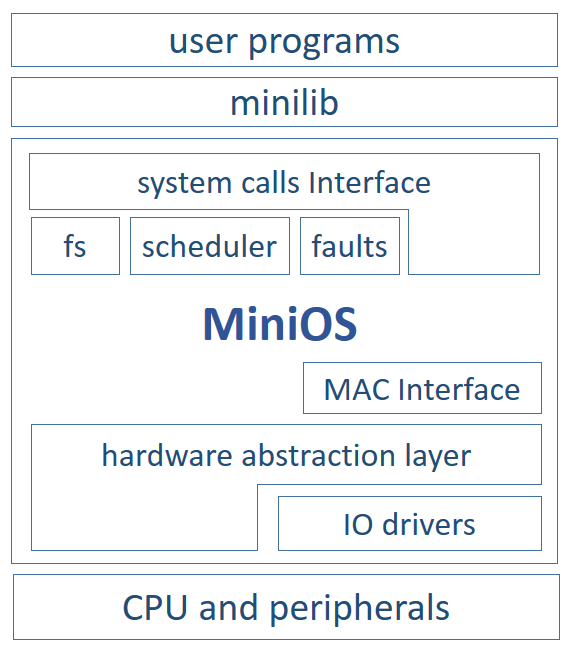}}
\subcaption{(c) 2015}
\end{minipage}%
\caption{MiniOS architecture in different years}
  \label{fig:minios-years}
\end{figure}

Reflecting on the experience to date, MiniOS has served well as an instructional system. Previous versions of the teaching platform (Figure~\ref{fig:minios-years}) were used as the laboratory component for Operating Systems in Fall 2013, 2014 and 2015, and as the laboratory component for Computer Architecture in Winter 2015 and 2016.

Figures~\ref{fig:minios-years}(a), ~\ref{fig:minios-years}(b), and ~\ref{fig:minios-years}(c) illustrate the evolution of MiniOS. Overall the delivery of the material went without problems, and in the process we gathered experiences.

\subsubsection{Book experience}

The role of the book seemed to have served its purpose. In particular, we think it seems to have allowed students to learn more complex OS topics (that would otherwise be out of scope). In other words, it seems that giving them all the basic details of OS construction allowed them to focus on more advanced features. 

The amount of details offered in its latest version allowed for students to complete assignments with little necessity to rely on other sources. In early offerings where the guide covered less material, students consistently indicated being frustrated for the lack of related external sources. Moreover, the amount of details seemed to have enabled students to complete projects. In Fall 2013 two of the five teams using the SAM4S board (two teams used different hardware) were not able to present working projects due problems of technical nature. In 2014, the number went down to one (that team used different hardware). In 2015, it went down to zero, and all teams used the SAM4S board. 

The system part of the guide assumes that students have had some experience working with the SAM4S board and Cortex-M architecture. When we delivered Operating Systems in 2015 this was not the case for every student, as some had taken Computer Architecture one year earlier (in 2014) using a different CPU architecture. Interestingly, the lack of previous Cortex-M experience did not seem to matter considerably. Three of the thirteen students attending labs did not have previous Cortex-M experience. Still, based on marks and personal appreciation, they performed similar to the rest of the class. In fact, one of them went to obtain the highest marks in labs. It is quite possible and reasonable that they dedicated more time to get on track with the new hardware platform and programming environment.

\subsubsection{Instruction and tutorial experience}

Based on previous experience teaching labs by building an OS for x86, MCU embedded hardware seemed to have allowed us to dedicate less instruction to present students with hardware details. Specially with the use of the guide, the required concepts previously introduced in Computer Architecture were just referenced and not re-introduced.

Tutorials were offered in a classroom once a week. Despite the material being covered in the guide, many times students needed further clarification. While some were able to finish labs with minimal or no consultation at all, others did not. Thus, it is recommended to have dedicated lab or tutorial sessions, where the lab instructor gives an oral presentation of the material. To gain insight into what is difficult and what is not, and what could use extra instruction, it is advised that the lab instructor solves the labs in advance. 

\subsubsection{Language experience}

Java is the language used to teach most courses at UNBC. This means that, for many students, Operating Systems was their first encounter with the C language. Among all the C-specific concepts, pointers and pointers to pointers demonstrated to be difficult to decipher. In fact, students consistently reported much of the struggle with assignments came from inexperience with C. To compensate for it, tutorials covered the use of pointers, callbacks, structures, organization of code in modules, use of header files, compiler attributes, among other relevant C specifics. Often students were not able to proceed further due to a specific language detail they were confused about or not knowledgeable of. While some students were prompt to ask, other were not. Those that did not ask reported spending a considerable amount of extra hours figuring out by themselves. Thus, students were encouraged to ask for language related doubts. In general, it is recommended that the lab instructor does not hesitate in assisting students with language problems.

\subsubsection{Debugger experience}

The presence of debugging facilities, albeit circumstantial, showed to be very important for solving the laboratory assignments. On occasions, the debugger was the difference between students completing an assignment or not. Often assistance was given in the form of debugging sessions. Sometimes because the lab instructor was unsure where the mistake was, and some other times because the debugger allowed for a demonstration of a concept that was otherwise not being understood from an oral explanation. Also, we have found that most of students' bugs are due not to wrong logic (they usually get it right), but due to a missing technical detail or a wrong assumption of technical nature. Debugging was very useful in finding those mistakes, as it enabled to verify step by step the details and assumptions of what is supposed to happen versus what is actually happening. This contrasts with typical remote debugging, which is a rather limited way of debugging (similar views are expressed by Holland et al \cite{OS161}). 

We also noted that in spite of previous debugging experience, in most cases, students lacked the debugging abilities to take advantage of the facilities available. In this regard, videos showing effective use of the debugger, as well as personal assistance, were provided. Interestingly, students seemed to have improved their bug-finding skills after only a couple short sessions of personal debugging assistance.  

\subsubsection{Hardware experience}

In the first offering of the course, the MCU platform had a neutral reception. This, however, has changed for the later two offerings of the course. Perhaps the guide played a role in that. For the latest offerings of both more than half of students showed interest and enthusiasm of working with hardware. 

Overall, boards behave well. On occasions, albeit not often, boards would simply fail to be recognized by Atmel Studio. Some times resetting the host computer or re-plugging the board would fix the problem, but other times the board would continue to fail and a replacement had to be given. So it is recommended to have extra boards in case they are needed.

Working with external peripherals can sometimes be a problem. There was a few incidents where Xbee modules and one board where burnt due to wrong wiring. Being computer science majors, a good number of students showed problems with wiring of external peripherals. For example, late in the course one student started having problems integrating an Xbee for his final project, and expressed that the OS course was (before the issues started) his favourite course in that semester. These problems repeated in a few occasions, and as a consequence, the latest version of MiniOS has stopped requiring any use of peripherals that are not expansion boards, since they simply plug into an expansion slot. Concretely, Xbee radio modules have been replaced by the REB233 board. At the time adding external support for PS/2 keyboards was also being considered, but had to be dismissed for the same reasons.

For final projects, students choose something of their interest, and often the wiring problem is impossible to avoid since they choose external hardware that requires wiring. Not only wiring, but finding the right piece of hardware have been consistently mentioned (in project reports) to be challenging. Our solution was to provide extra assistance. In some cases the assistance included videos showing how to wire the peripherals, and in other occasions sitting down with them and explaining the wiring in detail.

\subsubsection{Drivers experience}

An important aspect of MiniOS is the integration of third-party open source firmware. This enabled seamless integration of a variety of different peripherals. With available drivers, the adding of hardware functionality to the system became a mechanical task. We found this to be good for student engagement, as driver availability is the main limitation in using external hardware in projects. In fact, during project proposals we advised students to check for driver availability before acquiring any peripheral. It is worth noting that by themselves, drivers are of little use as their use is difficult to figure out from documentation. The device driver guide played an important role in simplifying it, and turning it into a mechanical task.

\subsection{Students' feedback}

In Fall 2014, students were asked to elaborate in what they thought were strengths and weaknesses of laboratories projects. These are some of the answers, organized by student number S1, S2, and so forth.

\begin{itemize}

\item S1 strengths: \emph{The assignments are very hands on and we get to see the things we discuss in class. The example programs show the functionality of the board.}

\item S1 weaknesses: \emph{there seems to be little documentation for the ASF. The coding can be hard to follow.}

\item S2: \emph{Assignments are a good way to see the complexity and challenge in dealing with the hardware level. They are nice in that you can access the hardware directly, and use the debugger that is provided to actually ``see'' the registers and the hex or binary values stored here, and how things are interacting.} 

\emph{However, that is also its main weakness. being that they are reasonable complex pieces of software it is very challenging to understand how all the components are interacting at times. [...] Along with the large amount of digging that needs to be done to understand the documentation, the other challenge is understanding C as I have not used it much to although things are similar it is still not Java. however it is kinda nice [illegible] to use something other than Java.}

\item S3 strengths: \emph{in my opinion, the programming assignments are a mixed bag. I think its a good way to show incrementally how each part of an OS is designed and implemented in practice. The assignment themselves strengthen the knowledge learned in class.}

\item S3 `weaknesses: \emph{the downside, however, is the language implemented. Its a minor point, but it is an issue with which I struggle. Until this course, I've never used/been/seen exposed to C. It just makes understanding and implementing ideas needlessly complex.}

\item S4: \emph{[...]It is too easy to get stuck on a simple task specially when the student is using a new language and development environment that is foreign.}

\emph{Being a Java university, the first assignment should a strictly C assignment. Designed to get an understanding of the differences from Java, and features required to use the Atmel libraries. This can be assigned day 1 of the term.}

\item S5 strengths: \emph{Got to see and develop an entire OS. Get to use a simple enough platform to feasibly develop all components. The interactive nature of the platform makes progress easy to see and rewarding. Tutorials are well written and provide detailed instructions. Code base is quite small and it is easy to hold entire system in your head.}

\item S5 weaknesses: \emph{Lack of documentation and online support for platform. Each assignment is very involved and requires a large time commitment.}

\item S6 strengths: \emph{The projects are fun, engaging and rewarding. Interacting with real hardware is great. The resources provided by the TA are complete and helpful}

\item S6 strengths: \emph{C is not something I am particularly familiar with. Atmel resources (documentation) are not always helpful. C language guides are not always applicable}

\end{itemize}

For fall 2015 substantial changes were made to the guide. In particular, it was made more self-contained, in the sense that it made little or no reference to external documentation. Additional missed necessary details dealing with the architecture were covered. Tutorial time was devoted for looking at specific C knowledge required to complete assignments. Certain embedded systems specific details not relevant to operating systems were removed. Students were again asked to elaborate in what they thought were strengths and weaknesses of laboratories projects. These are some of the answers (here we group them).

\begin{itemize}

\item Strengths:

\begin{itemize}

\item \emph{Physical depiction of what we're doing (interactive buttons \& oled screen)}

\item \emph{Overall great assignment layout. I don't understand why people need extensions!}

\item \emph{Everyone loves bonus questions!}

\item \emph{Nice to have many examples/viewpoints}

\item \emph{I've heard from many that they're having trouble with their board. Not me personally though.}

\item \emph{Great examples to draw from.}

\item \emph{Much prior use of board/software}

\item \emph{enforces understanding of: }

\begin{itemize}

\item \emph{Interrupt}
     
\item \emph{HAL}
     
\item \emph{better code}
     
\item \emph{better structure}
\end{itemize}
\item \emph{FUN!}

\item \emph{The documentation provided is top notch. It makes the world of a difference having lab documents and driver documents written in PLAIN LANGUAGE which speeds up learning.}

\item  \emph{Unified system. The SAM4S is easy to work with, we have been using it for a few years, so students aren't totally new to developing for it.}

\item \emph{Software support. It's undeniable Atmel Studio is useful in learning how to code for embedded systems. Visually stepping through code and viewing memory being modified helps intuition. As well as debugger.}

\item \emph{Relevant work, simplistic design. it's easy to develop on ARM and learn the ropes. Jumping to x86 would be more difficult. ARM is also very popular and won't be going away soon.}

\item \emph{Assignments are relevant to course. It's easier to break down OS concepts and learn how to code them [illegible]. it solidifies the ideas and makes classes easier to understand.}

\end{itemize}

\item Weaknesses:

\begin{itemize}

\item \emph{lack of any useful documentation about the SAM4S online }

\item \emph{C is a bit tough when no taught any prior C (pointers, memory is odd) (still not so hard)}

\item \emph{Atmel Studio 6.0, 6.2 a bit finicky and error prone (better with 7.0 now!)}

\item \emph{**For me* Many others would disagree:}                                 

\begin{itemize}
 
 \item \emph{not so hard enough sometimes}                                          

\item \emph{would like to build some driver from scratch (camera, touchscree, etc)}

\end{itemize}

\item \emph{Assignment are long. A lot of time is required to complete. Due to bugs it can sometimes take more than a week. Two weeks is usually required.}

\item \emph{Atmel Studio is buggy, it leads to a lot of wasted time messing around. }

\item \emph{It can sometimes be difficult to tie into classroom lectures. It would almost help to focus lecture on how ARM systems can have an OS build on them to have more insight.}

\end{itemize}

\end{itemize}

\subsection{Sample Final Projects}

\begin{figure}[ht]
\begin{minipage}{.5\textwidth}
\centerline{\includegraphics[scale=.5]{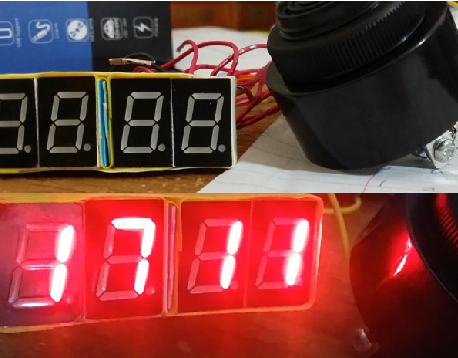}}
\subcaption{(a) 2013}
\end{minipage}%
\begin{minipage}{.5\textwidth}
\centerline{\includegraphics[scale=.6]{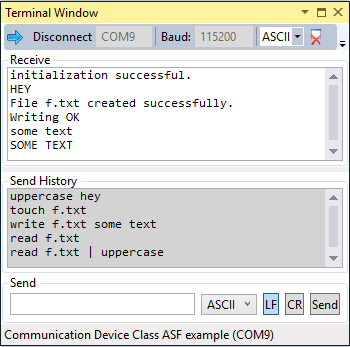}}
\subcaption{(a) 2013}
\end{minipage}%
\caption{Sample student projects}
\label{fig:student-projects}
\end{figure}

At the end of the semester we held, for both courses, a final project presentation, where students had to give a short presentation and demonstration of their projects. Figure~\ref{fig:student-projects} shows two sample projects. Figure~\ref{fig:student-projects}(a) shows a clock alarm project that runs MiniOS. Figure~\ref{fig:student-projects}(b) shows a project named \emph{Pinto pipes}, a rudimentary command line interpreter that supports redirection. Other projects include \emph{Ascii at a distance} (a communication API for wireless devices), \emph{SOS (simple operating system)}, \emph{Remote sensor drone} (a remote rover with sensing capabilities), thread signalling for MiniOS, and \emph{System Security} (secure user account management).  

\subsection{Laboratory and Project Grades}

Laboratory grading had two components. A programming component that accounted for around 80\% of the whole laboratory assignment grade, and a writing component accounting for 20\%. So, to get full marks students required not only working code, but also an understanding of the topic discussed in each lab. Non-working code was highly penalized with two thirds of the complete value for a given task (lab assignments were divided into tasks). Projects' grades were dependent on the difficulty and quality of the project, the amount of help asked for, and how it related to OS concepts (e.g. extending the system is preferred over using it as a tool). 

\begin{figure}[ht]
\begin{minipage}{.5\textwidth}
\centerline{\includegraphics[scale=.45]{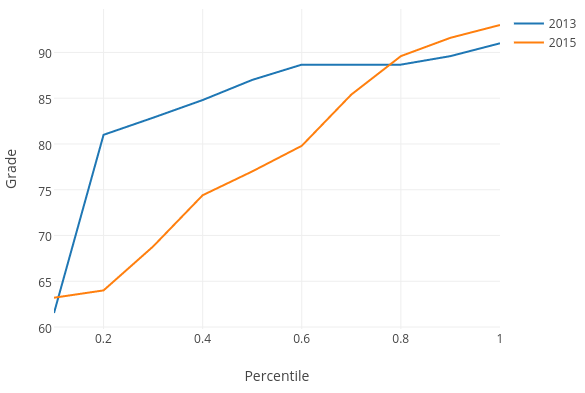}}
\subcaption{(a) Lab assignments}
\end{minipage}%
\begin{minipage}{.5\textwidth}
\centerline{\includegraphics[scale=.45]{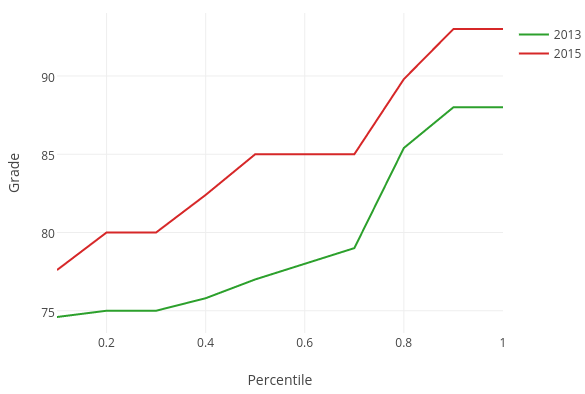}}
\subcaption{(a) Projects}
\end{minipage}%
\caption{Student Grades}
\label{fig:grades}
\end{figure}

Laboratory and project grades are available for 2013 and 2015 (in 2014 a different lab instructor taught labs) with a total of 17 and 13 attendants, respectively. Figure~\ref{fig:grades}a shows distribution of laboratory grades, while Figure~\ref{fig:grades}b shows distribution of project grades. 

The results show an adequate level of proficiency from students in both years. Incidentally, in 2015 marking was purposefully made more strict than in 2013 for both assignments and projects. That decision reflects in lab assignment's media dropping from 87\% in 2013 to 77\% in 2015. Interestingly, in 2015 project marks were higher, in spite of the harsher marking. The media went from 77\% in 2013 to 85\% in 2015. Moreover, 2015 was the only year when all students used the SAM4S board in their projects. Improved project marks may be due to (a) better student engagement in 2015 than in 2013 (we noticed students seemed more dedicated to their projects in 2015); and (b) a more developed guide that translated in better understanding of OS concepts and their implementation, which in turn resulted in projects of better quality, more complex, and more related to OS concepts.

\subsection{MiniOS as a prototyping research platform}

\begin{figure}[ht]
\centerline{\includegraphics[scale=.55]{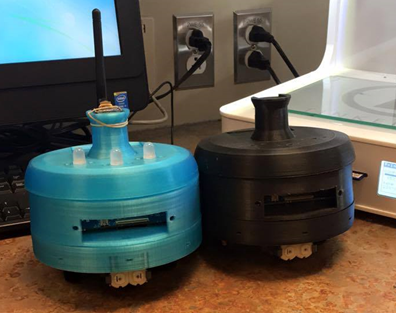}}
\caption{Multi-robot platform running MiniOS}
\label{fig:ruperts}
\end{figure}

The amount of functionality built in, together with the ease of hardware integration, made MiniOS a good alternative for embedded systems prototyping. In particular, applications have access to OS facilities, as well as straightforward sensor and actuator integration. Prototyping platforms with similar capabilities are Microsoft's .NET gadgeteer \cite{NetGadgeteer} and mbed \cite{mbed}. As example, consider Figure~\ref{fig:ruperts}, which shows two mobile robots part of an experimental multi-robot platform based on MiniOS. 

Likewise, given that MiniOS' inner working are well documented and are comparatively simpler that other systems, it could serve well as systems research platform where system designs can be tried in relatively small time frame.

Up to this point, we have discussed our appreciation of the teaching experience. The next section discusses feedback received from students.

%%%%%%%%%%%%%%%%%%%%%%%%%%%%%%%%%%%%%%%%%%%%%%%%%%%%%%%%%%
%%%%%     				S E C T I O N			      %%%%
%%%%%%%%%%%%%%%%%%%%%%%%%%%%%%%%%%%%%%%%%%%%%%%%%%%%%%%%%%
\section{Concluding Remarks}

\subsection{Conclusions}
The main contribution of this thesis is MiniOS, an instructional platform for the delivery of operating systems laboratories. MiniOS follows on the steps of instructional systems that attempt to deal with complexity by lowering the code volume. We go further and identify the target hardware platform as an additional source of complexity that can also be account for. The result is a MCU instructional operating system, and to our knowledge, the first one of its type. In addition, the platform includes a step-by-step guide to its construction whose purpose is to offer all the necessary details for the construction of the system. 

The platform was used in three different occasions to deliver laboratory assignments for an introductory course in operating systems. Student feedback was overall favourable. We presented this feedback together with other experiences.

MiniOS cannot---and is not intended to---replace more traditional desktop-centered 
design approaches; instead it serves as an alternative. For example, if a course has as objective to provide students with experience in topics related to the MMU (e.g. virtual memory), or to teach OSes as they are built in industry, the MiniOS approach is a poor fit. 

University laboratories are not the only place where MiniOS has found use. Given the amount of built-in functionality and the ease with which hardware can be integrated, it can, and has, been used as a rapid embedded prototyping platform. Likewise, being well-documented, it serves as a systems research test bed.
Currently the only existing ports is the SAM4S Xplained Pro board, but there is no reason that would not allow MiniOS to be ported to other MCU platforms. In fact, it was designed to be ported.

\subsection{Future Work}
For future work, we would like to explore the possibility of porting the instructional platform to other MCU platforms like Arduino Zero, a Cortex-M0 based Arduino. The integration of bare-metal drivers as part of MiniOS might work well with the plethora of available Arduino code online. 
Also, since there is existing infrastructure for wireless connectivity, it is possible to add internet connectivity and customize MiniOS to work as an IoT OS.

Finally, there is space for further experimental exploration of some of the observations made during the use of the platform in the past three years. For instance, the authors would like to test whether a smaller MCU platform does enable learning of the same concepts while requiring less instruction. Also, the impact that using the debugger has in explaining of technical concepts; among other observations made in this work.

% Appendix
\appendix
%\section*{APPENDIX}
%\setcounter{section}{1}
\appendixhead{ZHOU}

% Acknowledgments
%\begin{acks}
%\end{acks}

% Bibliography
\bibliographystyle{ACM-Reference-Format-Journals}
\bibliography{acmsmall-sample-bibfile}


\begin{thebibliography}{00}

%%% the Web address.

\ifx \showCODEN    \undefined \def \showCODEN     #1{\unskip}     \fi
\ifx \showDOI      \undefined \def \showDOI       #1{{\tt DOI:}\penalty0{#1}\ }
  \fi
\ifx \showISBNx    \undefined \def \showISBNx     #1{\unskip}     \fi
\ifx \showISBNxiii \undefined \def \showISBNxiii  #1{\unskip}     \fi
\ifx \showISSN     \undefined \def \showISSN      #1{\unskip}     \fi
\ifx \showLCCN     \undefined \def \showLCCN      #1{\unskip}     \fi
\ifx \shownote     \undefined \def \shownote      #1{#1}          \fi
\ifx \showarticletitle \undefined \def \showarticletitle #1{#1}   \fi
\ifx \showURL      \undefined \def \showURL       #1{#1}          \fi

\bibitem[\protect\citeauthoryear{Anderson and Nguyen}{Anderson and
  Nguyen}{2005}]%
        {Survey}
{Charles~L. Anderson} {and} {Minh Nguyen}. 2005.
\newblock \showarticletitle{A Survey of Contemporary Instructional Operating
  Systems for Use in Undergraduate Courses}.
\newblock {\em J. Comput. Sci. Coll.\/} {21}, 1 (Oct. 2005), 183--190.
\newblock
\showISSN{1937-4771}
\showURL{%
\url{http://dl.acm.org/citation.cfm?id=1088791.1088822}}


\bibitem[\protect\citeauthoryear{ARM}{ARM}{2016}]%
        {mbed}
{ARM}. 2016.
\newblock mbed.
\newblock   (2016).
\newblock
\showURL{%
\url{https://www.mbed.com/en/}}


\bibitem[\protect\citeauthoryear{Atkin and Sirer}{Atkin and Sirer}{2002}]%
        {portos}
{Benjamin Atkin} {and} {Emin~G\"{u}n Sirer}. 2002.
\newblock \showarticletitle{PortOS: An Educational Operating System for the
  Post-PC Environment}. In {\em Proceedings of the 33rd SIGCSE Technical
  Symposium on Computer Science Education} {\em (SIGCSE '02)}. ACM, New York,
  NY, USA, 116--120.
\newblock
\showISBNx{1-58113-473-8}
\showDOI{%
\url{http://dx.doi.org/10.1145/563340.563384}}


\bibitem[\protect\citeauthoryear{Ben-Ari}{Ben-Ari}{2016}]%
        {benari}
{Mordechai~(Moti) Ben-Ari}. 2016.
\newblock \showarticletitle{In Defense of Programming}.
\newblock {\em ACM Inroads\/} {7}, 1 (Feb. 2016), 44--46.
\newblock
\showISSN{2153-2184}
\showDOI{%
\url{http://dx.doi.org/10.1145/2827858}}


\bibitem[\protect\citeauthoryear{Black}{Black}{2009}]%
        {black}
{Michael~D. Black}. 2009.
\newblock \showarticletitle{Build an Operating System from Scratch: A Project
  for an Introductory Operating Systems Course}.
\newblock {\em SIGCSE Bull.\/} {41}, 1 (March 2009), 448--452.
\newblock
\showISSN{0097-8418}
\showDOI{%
\url{http://dx.doi.org/10.1145/1539024.1509022}}


\bibitem[\protect\citeauthoryear{Chadwick}{Chadwick}{2012}]%
        {cambridge}
{Alex Chadwick}. 2012.
\newblock Baking PI: Operating Systems Development.
\newblock   (2012).
\newblock
\showURL{%
\url{https://www.cl.cam.ac.uk/projects/raspberrypi/tutorials/os/}}


\bibitem[\protect\citeauthoryear{ChaN}{ChaN}{nown}]%
        {fatfs}
{ChaN}. Unknown.
\newblock FatFs - Generic FAT File System Module.
\newblock   (Unknown).
\newblock
\showURL{%
\url{http://elm-chan.org/fsw/ff/00index_e.html}}


\bibitem[\protect\citeauthoryear{Christopher, Procter, and
  Anderson}{Christopher et~al\mbox{.}}{1993}]%
        {nachos}
{Wayne~A. Christopher}, {Steven~J. Procter}, {and} {Thomas~E. Anderson}. 1993.
\newblock \showarticletitle{The Nachos Instructional Operating System}. In {\em
  Proceedings of the USENIX Winter 1993 Conference Proceedings on USENIX Winter
  1993 Conference Proceedings} {\em (USENIX'93)}. USENIX Association, Berkeley,
  CA, USA, 4--4.
\newblock
\showURL{%
\url{http://dl.acm.org/citation.cfm?id=1267303.1267307}}


\bibitem[\protect\citeauthoryear{Comer}{Comer}{1984}]%
        {comer}
{Douglas Comer}. 1984.
\newblock {\em Operating System Design: The XINU Approach}.
\newblock Prentice-Hall, Inc., Upper Saddle River, NJ, USA.
\newblock
\showISBNx{0-13-637539-1}


\bibitem[\protect\citeauthoryear{Corliss and Melara}{Corliss and
  Melara}{2011}]%
        {vireos}
{Marc~L. Corliss} {and} {Marcela Melara}. 2011.
\newblock \showarticletitle{VIREOS: An Integrated, Bottom-up, Educational
  Operating Systems Project with FPGA Support}. In {\em Proceedings of the 42Nd
  ACM Technical Symposium on Computer Science Education} {\em (SIGCSE '11)}.
  ACM, New York, NY, USA, 39--44.
\newblock
\showISBNx{978-1-4503-0500-6}
\showDOI{%
\url{http://dx.doi.org/10.1145/1953163.1953179}}


\bibitem[\protect\citeauthoryear{Fankhauser, Conrad, Zitzler, and
  Plattner}{Fankhauser et~al\mbox{.}}{2000}]%
        {topsy}
{George Fankhauser}, {Christian Conrad}, {Eckart Zitzler}, {and} {Bernhard
  Plattner}. 2000.
\newblock {\em Topsy -- A Teachable Operating System}.
\newblock
\showURL{%
\url{http://www.tik.ee.ethz.ch/~topsy/Book/Topsy_1.1.pdf}}


\bibitem[\protect\citeauthoryear{Goldweber, Davoli, and Morsiani}{Goldweber
  et~al\mbox{.}}{2005}]%
        {kaya}
{Michael Goldweber}, {Renzo Davoli}, {and} {Mauro Morsiani}. 2005.
\newblock \showarticletitle{The Kaya OS Project and the uMPS Hardware
  Emulator}.
\newblock {\em SIGCSE Bull.\/} {37}, 3 (June 2005), 49--53.
\newblock
\showISSN{0097-8418}
\showDOI{%
\url{http://dx.doi.org/10.1145/1151954.1067462}}


\bibitem[\protect\citeauthoryear{Guzdial}{Guzdial}{2015}]%
        {guzdial}
{Mark Guzdial}. 2015.
\newblock \showarticletitle{What's the Best Way to Teach Computer Science to
  Beginners?}
\newblock {\em Commun. ACM\/} {58}, 2 (Jan. 2015), 12--13.
\newblock
\showISSN{0001-0782}
\showDOI{%
\url{http://dx.doi.org/10.1145/2714488}}


\bibitem[\protect\citeauthoryear{Holland, Lim, and Seltzer}{Holland
  et~al\mbox{.}}{2002}]%
        {OS161}
{David~A. Holland}, {Ada~T. Lim}, {and} {Margo~I. Seltzer}. 2002.
\newblock \showarticletitle{A New Instructional Operating System}. In {\em
  Proceedings of the 33rd SIGCSE Technical Symposium on Computer Science
  Education} {\em (SIGCSE '02)}. ACM, New York, NY, USA, 111--115.
\newblock
\showISBNx{1-58113-473-8}
\showDOI{%
\url{http://dx.doi.org/10.1145/563340.563383}}


\bibitem[\protect\citeauthoryear{Hovemeyer, Hollingsworth, and
  Bhattacharjee}{Hovemeyer et~al\mbox{.}}{2004}]%
        {GeekOS}
{David Hovemeyer}, {Jeffrey~K. Hollingsworth}, {and} {Bobby Bhattacharjee}.
  2004.
\newblock \showarticletitle{Running on the Bare Metal with GeekOS}.
\newblock {\em SIGCSE Bull.\/} {36}, 1 (March 2004), 315--319.
\newblock
\showISSN{0097-8418}
\showDOI{%
\url{http://dx.doi.org/10.1145/1028174.971411}}


\bibitem[\protect\citeauthoryear{Levis, Patel, Culler, and Shenker}{Levis
  et~al\mbox{.}}{2004}]%
        {trickle}
{Philip Levis}, {Neil Patel}, {David Culler}, {and} {Scott Shenker}. 2004.
\newblock \showarticletitle{Trickle: A Self-regulating Algorithm for Code
  Propagation and Maintenance in Wireless Sensor Networks}. In {\em Proceedings
  of the 1st Conference on Symposium on Networked Systems Design and
  Implementation - Volume 1} {\em (NSDI'04)}. USENIX Association, Berkeley, CA,
  USA, 2--2.
\newblock
\showURL{%
\url{http://dl.acm.org/citation.cfm?id=1251175.1251177}}


\bibitem[\protect\citeauthoryear{Liu, Chen, and Gong}{Liu
  et~al\mbox{.}}{2007}]%
        {babyos}
{Haifeng Liu}, {Xianglan Chen}, {and} {Yuchang Gong}. 2007.
\newblock \showarticletitle{BabyOS: A Fresh Start}. In {\em Proceedings of the
  38th SIGCSE Technical Symposium on Computer Science Education} {\em (SIGCSE
  '07)}. ACM, New York, NY, USA, 566--570.
\newblock
\showISBNx{1-59593-361-1}
\showDOI{%
\url{http://dx.doi.org/10.1145/1227310.1227499}}


\bibitem[\protect\citeauthoryear{Ltd}{Ltd}{2016}]%
        {freertos}
{Real Time~Engineers Ltd}. 2016.
\newblock FreeRTOS.
\newblock   (2016).
\newblock
\showURL{%
\url{http://www.freertos.org/}}


\bibitem[\protect\citeauthoryear{Moseley and Marks}{Moseley and Marks}{2006}]%
        {tarpit}
{Ben Moseley} {and} {Peter Marks}. 2006.
\newblock \showarticletitle{Out of the Tar Pit}. In {\em SOFTWARE PRACTICE
  ADVANCEMENT (SPA)}.
\newblock


\bibitem[\protect\citeauthoryear{Pfaff, Romano, and Back}{Pfaff
  et~al\mbox{.}}{2009}]%
        {pintos}
{Ben Pfaff}, {Anthony Romano}, {and} {Godmar Back}. 2009.
\newblock \showarticletitle{The Pintos Instructional Operating System Kernel}.
  In {\em Proceedings of the 40th ACM Technical Symposium on Computer Science
  Education} {\em (SIGCSE '09)}. ACM, New York, NY, USA, 453--457.
\newblock
\showISBNx{978-1-60558-183-5}
\showDOI{%
\url{http://dx.doi.org/10.1145/1508865.1509023}}


\bibitem[\protect\citeauthoryear{Pinto, Nobile, Mamani, J�nior, Luz, and
  Monaco}{Pinto et~al\mbox{.}}{2013}]%
        {tempos}
{Ren�~S. Pinto}, {Pedro Nobile}, {Edwin Mamani}, {Louren�o~P. J�nior},
  {Helder~J.F. Luz}, {and} {Francisco~J. Monaco}. 2013.
\newblock \showarticletitle{Operating System from the Scratch: A Problem-based
  Learning Approach for the Emerging Demands on \{OS\} Development}.
\newblock {\em Procedia Computer Science\/} {18}, 0 (2013), 2472 -- 2481.
\newblock
\showISSN{1877-0509}
\showDOI{%
\url{http://dx.doi.org/10.1016/j.procs.2013.05.424}}
\newblock
\shownote{2013 International Conference on Computational Science.}


\bibitem[\protect\citeauthoryear{Svec}{Svec}{2016}]%
        {FreeRTOSLOC}
{Christopher Svec}. 2016.
\newblock FreeRTOS {@ONLINE}.
\newblock   (2016).
\newblock
\showURL{%
\url{http://www.aosabook.org/en/freertos.html}}


\bibitem[\protect\citeauthoryear{TanenBaum and Woodhull}{TanenBaum and
  Woodhull}{2006}]%
        {minix}
{Andrew~S. TanenBaum} {and} {Albert~S. Woodhull}. 2006.
\newblock {\em Operating System Design and Implementation\/} (third ed.).
\newblock Pearson.
\newblock


\bibitem[\protect\citeauthoryear{Tang}{Tang}{2015}]%
        {scheme}
{Jonathan Tang}. 2015.
\newblock Write Yourself a Scheme in 48 Hours.
\newblock   (2015).
\newblock
\showURL{%
\url{https://en.wikibooks.org/wiki/Write_Yourself_a_Scheme_in_48_Hours}}


\bibitem[\protect\citeauthoryear{Unknown}{Unknown}{2013}]%
        {xinuwiki}
{Unknown}. 2013.
\newblock {\em Embedded Xinu}.
\newblock
\showURL{%
\url{http://xinu.mscs.mu.edu/Teaching_With_Xinu}}


\bibitem[\protect\citeauthoryear{Unknown}{Unknown}{2015}]%
        {AndroidLOC}
{Unknown}. 2015.
\newblock XDA-Developers:Android.
\newblock   (Jan. 2015).
\newblock
\showURL{%
\url{http://forum.xda-developers.com/wiki/XDA-Developers:Android}}


\bibitem[\protect\citeauthoryear{Villar, Scott, Hodges, Hammil, and
  Miller}{Villar et~al\mbox{.}}{2012}]%
        {NetGadgeteer}
{Nicolas Villar}, {James Scott}, {Steve Hodges}, {Kerry Hammil}, {and} {Colin
  Miller}. 2012.
\newblock \showarticletitle{.NET Gadgeteer: A Platform for Custom Devices}. In
  {\em Proceedings of the 10th International Conference on Pervasive Computing}
  {\em (Pervasive'12)}. Springer-Verlag, Berlin, Heidelberg, 216--233.
\newblock
\showISBNx{978-3-642-31204-5}
\showDOI{%
\url{http://dx.doi.org/10.1007/978-3-642-31205-2_14}}


\end{thebibliography}
                             % Sample .bib file with references that match those in
                             % the 'Specifications Document (V1.5)' as well containing
                             % 'legacy' bibs and bibs with 'alternate codings'.
                             % Gerry Murray - March 2012

% History dates

% Electronic Appendix
\elecappendix

\medskip

\section{Excerpt from Executing Applications Chapter}
\label{sec:appendix-apps}
\noindent
\includegraphics[
    page=1,
    width=\textwidth,
    height=\textheight,
    keepaspectratio
]{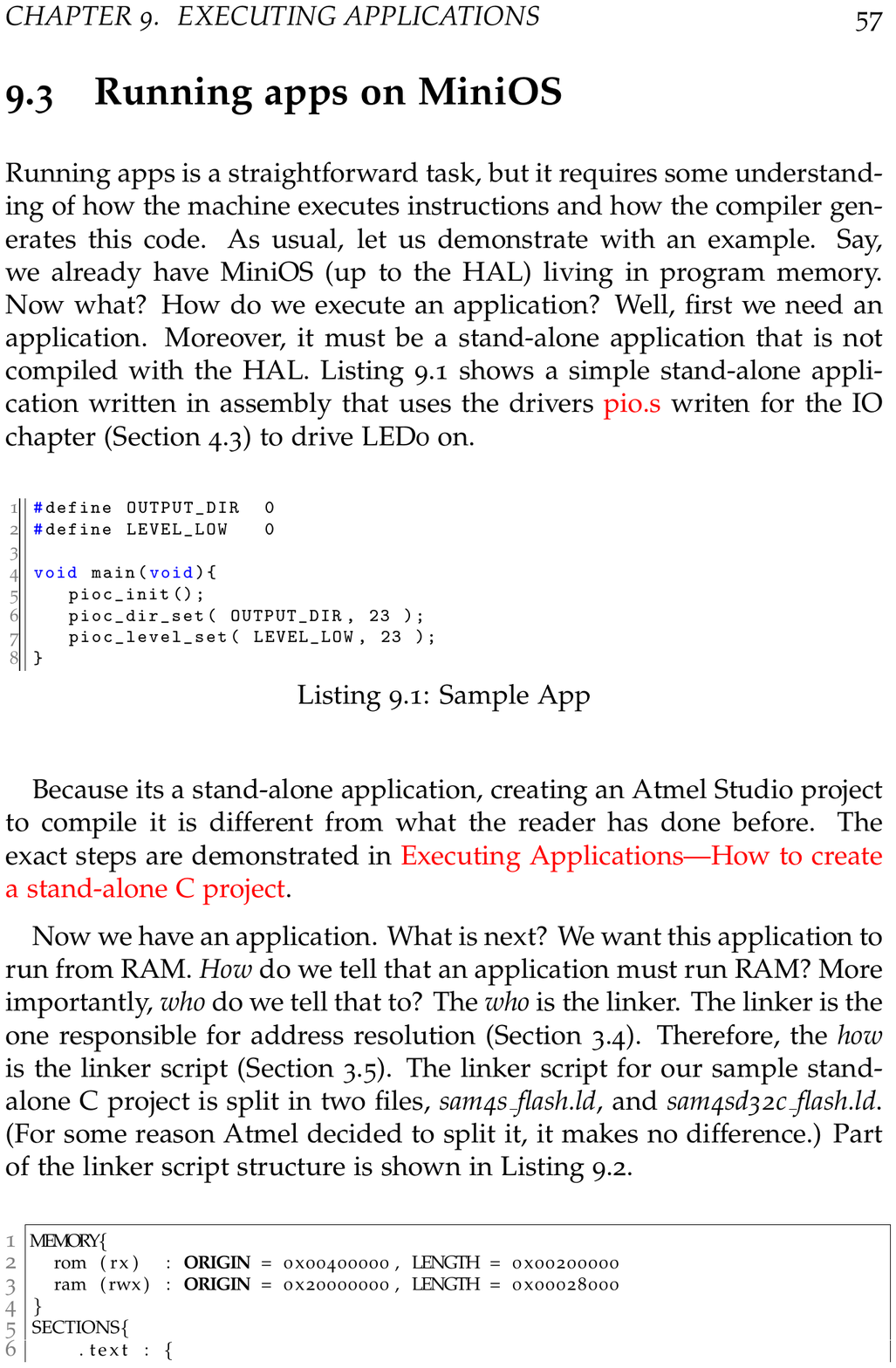}

\section{Excerpt from System Calls Chapter}
\label{sec:appendix-syscalls}
\noindent
\includegraphics[
    page=2,
    width=\textwidth,
    height=\textheight,
    keepaspectratio
]{./appendix.pdf}

\section{Excerpt from Device Driver's Section}
\label{sec:appendix-drivers}
\noindent
\includegraphics[
    page=3,
    width=\textwidth,
    height=\textheight,
    keepaspectratio
]{./appendix.pdf}

\end{document}